\documentclass[a4paper, 12pt, oneside]{article}

\usepackage[T1]{fontenc}
\usepackage [english]{babel}
\usepackage{natbib} 
\usepackage{authblk}

\usepackage{amsmath,amssymb}
\usepackage{booktabs}
\usepackage{tabularx}
\usepackage{latexsym}
\usepackage{amsfonts}
\usepackage{mathtools} 
\usepackage{bm}
\usepackage{dsfont}
\usepackage{bbm}

\usepackage{amsthm}
\usepackage{amscd}
\usepackage[inline]{enumitem}

\usepackage{multirow}
\usepackage{graphicx,graphics,rotating}
\usepackage[graphicx]{realboxes} 
\usepackage{epstopdf} 
\usepackage[font=scriptsize,labelfont=bf,labelsep=period,textfont=it]{caption}

\usepackage{mathrsfs}
\usepackage[a4paper,top=1in,bottom=1in,left=1in,right=1in,height rounded,bindingoffset=0cm]{geometry}

\usepackage{etoolbox}
\usepackage{xpatch}  
\usepackage[gen]{eurosym}
\usepackage{lscape}
\usepackage[toc,page]{appendix}

\usepackage{sgame}

\usepackage{setspace}
\usepackage[font={small,sl}]{caption}
\usepackage{booktabs,caption} 
\usepackage[para,online,flushleft]{threeparttable}	

\usepackage{blindtext}
\usepackage{fancyhdr}
\usepackage{setspace}
\usepackage[explicit]{titlesec}
\usepackage[normalem]{ulem}
\usepackage{lipsum}
\usepackage{mwe}  
\usepackage{verbatim}
\usepackage{booktabs,array}  
\usepackage{listings}
\usepackage{subfig}
\usepackage[colorlinks=true,citecolor=blue,linkcolor = blue,urlcolor  = blue]{hyperref}
\usepackage{xcolor}
\usepackage{booktabs}
\usepackage[hang,flushmargin]{footmisc}

\usepackage{mathptmx}

\makeatletter

\def\citeapos#1{\citeauthor{#1}'s (\citeyear{#1})}

\title{\Large Gender Segregation: Analysis across Sectoral-Dominance in the UK Labour Market}
\author[1,2]{Riccardo Leoncini} 
\author[1]{Mariele Macaluso\footnote{(Corresponding author: mariele.macaluso2@unibo.it). We are grateful to David Card for the helpful direction, especially related to the literature. We thank Emma Duchini, Marco Francesconi, Giovanni Guidetti and Giulio Pedrini for their helpful comments. We also thank conference participants at the AASLE 2022 Conference, 4$^{th}$ International Conference of European Studies, Summer School in Labor Economics at Barcellona School of Economics, 2023 Petralia Workshop. Our thanks go also to the editor and two anonymous reviewers for their detailed and useful comments. Annalivia Polselli acknowledges financial support from the South East Network for Social Sciences (SeNSS) as part of the Doctoral Training Partnership (award ES/P00072X/1/2128216).}} 
\author[3]{Annalivia Polselli}

\affil[1]{\small Department of Legal Studies, Alma Mater Studiorum, University of Bologna.}
\affil[2]{\small IRCrES, National Research Council, Milan.}
\affil[3]{\small Department of Economics, University of Essex.}
\newcommand{\distas}[1]{\mathbin{\overset{#1}{\kern\z@\sim}}}%

\newcommand{\E}{\mathbb{E}}	
\newcommand{\X}{\mathbf{X}}
\newcommand{\bbeta}{\bm{\beta}}

\vspace{-4mm}
	
\date{\today}

\begin{document}
	\maketitle
	\vspace{-10mm}
\begin{abstract}
	\noindent 
This paper aims to evaluate how changing patterns of sectoral gender segregation play a role in accounting for women's employment contracts and wages in the UK between 2005 and 2020. We then study wage differentials in gender-specific dominated sectors. 
We found that the propensity of women to be distributed differently across sectors is a major factor contributing to explaining the differences in wages and contract opportunities. Hence, the disproportion of women in female-dominated sectors implies contractual features and lower wages typical of that sector, on average, for all workers. This difference is primarily explained by ``persistent discriminatory constraints'', while human capital-related characteristics play a minor role.  However, wage differentials would shrink if workers had the same potential and residual wages as men in male-dominated sectors. Moreover, this does not happen at the top of the wage distribution, where wage differentials among women working in female-dominated sectors are always more pronounced than those of men.

\def\baselinestretch{1}

\setlength{\topmargin}{-0.6in} \setlength{\textheight}{9.2in}
\setlength{\oddsidemargin}{-0.1in}
\setlength{\evensidemargin}{-0.1in} \setlength{\textwidth}{6.7in}


\end{abstract}

\doublespacing
\newpage
\section{Introduction}
Over the last two decades, there has been a large increase in the labour market participation of women in several OECD countries \citep{oecd2023}. However, the rise in female employment rates has primarily interested sectors where women are already over-represented, such as health care, food and accommodation, and service activities \citep{OECD2020security,Eurofound2021}. 
Between 2005 and 2020, the share of women in total employment in the United Kingdom (UK) exceeded 70\% in sectors such as education, health, and households as employers\footnote{The sector of \textit{Households as Employers} includes the activities of domestic personnel (e.g., maids, cooks, waiters, gardeners, chauffeurs, caretakers, babysitters, etc.) whose product is consumed by the employing household (ONS: \href{http://www.siccodesupport.co.uk/sic-division.php?division=97}{http://www.siccodesupport.co.uk/sic-division.php?division=97}).}, while it was below 30\% in sectors like agriculture, mining and quarrying, manufacturing, construction, and transport  (Table~\ref{tab:sectors1}). %
 
To support equal treatment of workers in the workplace and improve gender diversity across industries, the UK adopted several reforms over the past decade, including the Equality Act 2010 (EA2010) that sets out several measures prohibiting, among others, gender discrimination in employment and pay.\footnote{According to the EA2010 and its related extensions (e.g. Regulations 2011 - Specific Duties and Public Authorities), a woman must not be discriminated against with respect to a man in a similar situation or due to a particular policy or working practice. For an updated map of the key policies regarding equality for and between women, see the UK Parliament website \href{https://www.parliament.uk/globalassets/documents/commons/scrutiny/gender_equality_policy_map_april_2020.pdf}{here}.} Although these policies have led to more balanced participation rates \citep{ons2022}, industrial segregation is still a major factor contributing to explaining the sorting of women across occupations and the labour market differentials, including employment and wage gap \citep{olsen2018,geo2022,irvine2022}.\footnote{Tables~\ref{tab:sic_soc_ml} and \ref{tab:sic_soc_fml} in the Appendix show that occupations in traditionally female-dominated sectors (e.g. education) are mainly over-represented by women, even in those occupations that are usually considered male-dominated (e.g. manager, director and senior officials, or skilled trades).} 
For instance, \citet{razzu2018} found that the reduction of women's representation in certain industrial sectors -- such as manufacturing and banking \& finance -- is responsible for the shifts in female employment after the 1990s, although the gap between women and men in terms of human capital characteristics has closed over time. In addition, female labour supply in in-person sectors has recently faced severe disruption due to the COVID-19 outbreak -- especially for young women, working mothers, and female immigrants \citep{czymara2020,opensociety2020,guardian2021}.\footnote{Many recent studies have defined COVID-19 pandemic as \emph{she-session}, showing that this crisis has significantly hit women with and without children especially in female-dominated sectors \citep{gupta2020,goldin2022}.}

 
This study uses the UK Labour Force Survey (LFS) quarterly data between 2005 and 2020 to investigate two questions: (i) how gender segregation across sectors contributes to explaining the type of employment contracts (i.e. part-time, permanent, remote work, number of weekly working hours) and hourly wages for women and men; (ii) how wage differentials in female- and male-dominated sectors differ between women and men in terms of observable and unobservable characteristics.


The first question is addressed through Propensity Score Matching (PSM) by estimating the average differences in labour market outcomes between workers in female- and male-dominated sectors with similar observed socio-demographic and working characteristics. 

To answer the second question, we first use the three-fold \citet{kitagawa1955}-\citet{blinder1973}-\citet{oaxaca1973} (KBO) decomposition to explore differences in wages due to human capital and productivity, or unexplained factors, or their simultaneous effect. We then estimate Mincer wage regression to explore the contribution of human capital and other observable skills. However, since there might be unobservable factors -- e.g., behavioural traits such as self-esteem, ambition, competitiveness and willingness to make risky career choices \citep{gneezy2003, gneezy2004,booth2009,Bertrand2011, saccardo2018} -- that may contribute to driving wage differentials, we retrieve predicted and residual wages from Mincer regression. We also conducted a counterfactual analysis to study how predicted and residual wages differ if workers (women in male- and female-dominated sectors and men in female-dominated sectors) had the same characteristics as men in male-dominated sectors.



Our main findings can be summarised in three points. First,  gender-based sectoral segregation matters in the disparity of contractual opportunities, even controlling for occupational composition. Workers in female-dominated sectors are more likely to be segregated into atypical contracts (part-time), to work fewer hours and less from home, and to earn less than their counterparts in male-dominated sectors. The penalty for men working in those sectors is even larger than for women.
Second, human capital and background characteristics play a minor role in explaining wage differentials. Instead, most of the difference is driven by persistent ``discriminatory constraints'' \citep{altonji1999}, such as barriers in the labour market for women due to the effects of discrimination (i.e. unequal pay for equally qualified workers) and unobserved differences in productivity and tastes.
Third, wage differentials in female- and male-dominated sectors would shrink if workers had the same potential and residual wages as men in male-dominated sectors. However, women in female-dominated sectors would always earn less than men in high-paid jobs due to the negative selection in the labour market.

While most of the literature focus on the role of gender segregation in explaining the gender wage gap by looking at occupational and job dimensions \citep{blackburn1993,watts1992,watts1995,watts1998,petrongolo2004,CortesPan2017,folke2020,scarborough2021},\footnote{These studies found a threefold explanation for gender differences: (i) women's preferences for more flexible and family-oriented contracts \citep{petrongolo2004,Bertrand2011,goldin2014,Bertrand2020,morchio2021} and less competitive and risky environments  \citep{gneezy2003,saccardo2018}; (ii) comparative advantage in terms of human capital and productivity \citep{petrongolo2004,petho2021}; (iii) discrimination \citep{petrongolo2004} and sexual harassment \citep{folke2020}.} our work is closely related to the scant literature on the role of gender segregation across sectors \citep{moir1979,kreimer2004,campos2016,kamerade2018,scarborough2021}. These papers highlight how gender division of labour is still embedded in sectors \citep{Carvalho2019}, which is considered a structural factor shaping the differential effects on labour markets caused by economic recessions  \citep{rubery2010,rubery2013,kamerade2018} and the business cycle \citep{hoynes2012,Perivier2014,doepke2016,razzu2020,pilatowska2022}. For instance, \citet{olivetti2014} emphasise the interplay between gender trends and the evolution of the industry structure. Therefore, an understanding of the sectoral composition of the workforce is necessary to assess the trajectory of male and female employment and wages \citep{moir1979}.\footnote{The gendered division of labour across sectors, mainly due to persistent social norms and stereotypes, influences how women self-select into different jobs and careers and bargain their contracts \citep{Card2016}, and is likely to distort preferences, labour market trajectories and future wages \citep{mumford2008,reuben2017,cortespan2020,jewell2020,card2021}.}
Our contribution focuses on three main points. First, an innovative feature of this study is that it goes beyond the standard segregation index to identify female- and male-dominated sectors by further classifying sectors into high- and low-segregated. This index allows us to measure the degree of imbalance and the intensity of differences in gender distribution of a sector over time. 
Second, our analysis is not limited to the results of the KBO that show which effect drives wage differentials. Instead, we disentangle the effect of differences in average characteristics of male and female employees and the effect of selection into different sectors of women and men on the gender wage differentials.
Third, since unobserved confounders may enter the decision to choose a specific (female or male) sector, we address this potential selection bias by calculating the predicted and residual wages from the Mincer regression. This approach is similar to the method used in the literature on migration to calculate individual potential earnings \citep{parey2017} and capture the part of earnings that is uncorrelated to observed skills \citep{gould2016,borjas2019}.\footnote{This literature highlights that immigrants could be positively/negatively selected based on both observed (e.g., higher levels of education) and unobserved determinants of labour market success (e.g. motivation, ambition and ability) that can enter into the decision to self-select into migration \citep{chiswick1978,chiswick1986,chiswick1999,borjas1987,bertoli2016}.}

Finally,  we extend the findings of the 1980s literature on the issue of ``comparable worth''\footnote{Comparable worth, or ``the women's issue of 1980'', was a wage-setting policy on a firm-by-firm basis proposed to reduce the gender gap in earnings. Accordingly, jobs within a firm with comparable worth should receive equal compensation \citep{ehrenberg1987}. This was based on job evaluation scores to compare jobs in different occupations \citep{madden1987}. However, this implementation was not without disagreements \citep[e.g.][]{usa1984,gleason1985,ferber1986,aaron1987,gerhart1991}.} \citep{treiman1981,maahs1985, bielby1986,aaron1987}. This literature found that the disproportion of women in female-dominated occupations is associated with lower pay in that occupation, on average, for all employees -- men and women \citep{treiman1981, killingsworth1987}. However, the negative effect on the wage of being in such jobs is more significant for men than women \citep{roos1981}, even after controlling for relevant worker and job characteristics, including industry effects \citep{johnson1984}. We find that these negative results -- i.e., worst job characteristics and wages for both men and women in female-dominated environments -- are confirmed even when looking at industrial sectors. However, we found a more pronounced wage differential among women than men in female-dominated sectors at the top of the wage distribution. 


The rest of the paper is structured as follows. Section~\ref{sec:conceptual} discusses the measures of gender sectoral dominance and segregation. Section~\ref{sec:data} describes the data and reports some descriptive analysis. Section~\ref{sec:methods} presents the empirical strategy. Section~\ref{sec:results} reports the estimated results. Section~\ref{sec:conclusion} concludes.

\section{Conceptual Framework}\label{sec:conceptual}
\subsection{Gender Sectoral Segregation Index}\label{sec:segregation}
Gender sectoral segregation arises when a disproportionate share of men or women exists in a sector of the economy, independent of the nature of the job allocation  \citep{watts1998}. In this section, we introduce the notion and measures of gender segregation in industrial sectors.

A sector is \emph{female-dominated (fml-dom)} if the share of women employed in that sector is higher than the share of men in the same sector; it is \emph{male-dominated (ml-dom)} if the share of men is higher than the share of women in the same sector. In formulae, the classification criterion for gender sectoral dominance is as follows:
\begin{equation}\label{eqn:frac}
\text{Sectoral Dominance} = 
\begin{cases}
\text{Female} & \text{ if } \frac{W_{jt}}{W_{t}} > \frac{M_{jt}}{M_{t}}\\
\text{Male }& \text{ otherwise } 
\end{cases}
\end{equation}
\noindent where $W_{jt}$ and $M_{jt}$ are respectively the total number of women and men employed in sector $j$ (SIC 1-digit) at time $t$;  $W_{t}$ and $M_{t}$ are respectively the total number of female and male workers at time $t$. 

The classification criterion defined in~\eqref{eqn:frac} uses the ``majority voting'' rule -- i.e., the group with the largest number of members (either male or female) represents the sector.\footnote{The denominators in~\eqref{eqn:frac} are not total employment (i.e., male plus female employees) but total employment by gender group, providing the overall share of women (or men) in a sector. The advantage of the criterion consists in avoiding the use of conservative thresholds of more than 60\% \citep{killingsworth1987} to classify a sector as female-dominated.} 

Based on classification criterion~\eqref{eqn:frac}, we define the Sectoral Segregation Index ($SSI^s_t$) as a measure of the degree of disproportion in the distributions of men and women in female- and male-dominated sectors at each time period. The index is based on the well-known Index of Dissimilarity,\footnote{The Index of Dissimilarity measures the degree of the disproportion in the distributions of two groups. It provides information on the proportion of the minority group that would have to be transferred to reach no segregation \citep{cortese1976, zoloth1976,watts1998}. \citet{duncan1955}'s seminal article provides the first systematic review of segregation indices. A more comprehensive approach to segregation measurement has been available since the 1980s  \citep{reardon2002}.} which is used in labour \citep{watts1998} and education economics \citep{zoloth1976,james1985} to study group composition and quantify the segregation among two groups \citep{cortese1976}. 

$SSI^s_t$ is calculated for the two gender dominated sectors ($SSI^{fml-dom}$ and $SSI^{ml-dom}$) as follows:
\begin{equation}\label{eqn:ssi}
SSI^s_{t} = \frac{1}{2}\sum_{j\in J_s} \bigg|\frac{W_{jt}}{W_{t}} - \frac{M_{jt}}{M_{t}} \bigg| \,\,\, \text{ for all } t \text{ and } s = \{ml-dom, fml-dom\}
\end{equation}
\noindent where $J_s$ is the set of sectors in male-dominated or female-dominated group. The index informs on time-varying group imbalance within gender-dominated sectors and ranges between 0 and 1.  Large values of $SSI^{fml-dom}$ (or $SSI^{ml-dom}$) flag large gender imbalance towards women (or men) and indicate the proportion of women (or men) that would have to either leave or enter each sector to avoid gender sectoral segregation. The value of the index remains unchanged when transferring workers between sectors within each gender group.\footnote{A similar interpretation is provided by \citet{zoloth1976} to describe the racial composition of schools within and across districts. In her setting, the Dissimilarity Index ($D$) remains unaffected by transferring students between schools within each group (minority and non-minority students). $D$ changes only by transferring students across the two groups.} 

Because the difference in the share of female and male employees can be extremely low in some sectors or high in others, we define an additional index that allows us to distinguish between high- and low-segregated sectors. Specifically, on average, sectors that display low (or high) segregation are classified as low (or highly) segregated sectors by ranking them from the least to the most segregated, based on $SSI^s_{t}$.


Our analysis follows the UK Standard Industrial Classification (SIC) at one digit, which is used for business establishments by type of economic activity. We consider the 19 sectors listed in Table~\ref{tab:sectors2} as female- or male-dominated sectors and by the degree of segregation (high and low).\footnote{For years before 2008, we used the correspondence between SIC 2003 and SIC 2007 sections. Sectors labelled as \emph{\mbox{O - Public} administration and defence} and \emph{U - Extra territorial} are removed from the sample due to the different nature of contracts and wages in their related jobs.}

\begin{figure}[ht!]
    \caption{Evolution of SSI index over fiscal year}\label{fig:line_id}
    \centering
    \subfloat{\includegraphics[scale=.85]{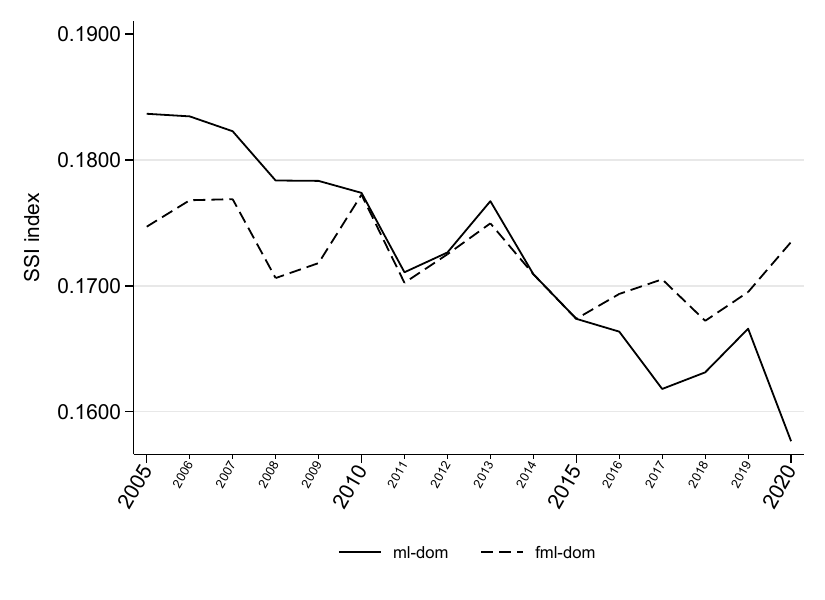}}\\
    \subfloat{\includegraphics[scale=.8]{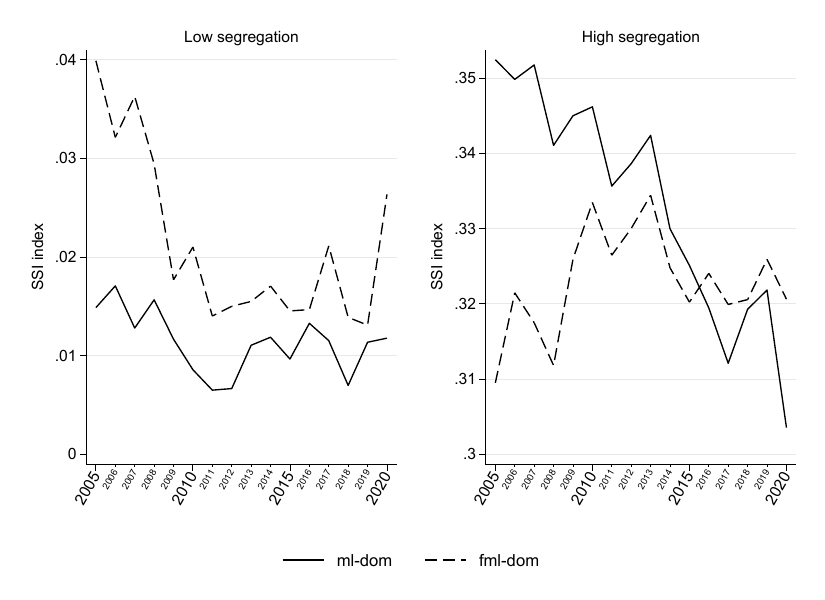}}\\
    \vspace{-1mm}
\end{figure}

Figure~\ref{fig:line_id} shows the evolution of $SSI^s_t$ over 2005-2020 in the UK economy (upper graph), in low-segregated (bottom left graph), and high-segregated sectors (bottom right graph). The index follows a downward trajectory in male-dominated sectors (from 0.182 in 2005 to 0.158 in 2020), meaning that sectoral segregation has decreased. Instead, the segregation remains stable at around 0.17 over time in female-dominated sectors. In low-segregated sectors, gender segregation is larger in female-dominated than male-dominated sectors (with mean values of 0.021 and 0.011, respectively). Instead, in high-segregated sectors, gender segregation is higher in male-dominated sectors. However, it decreased over time (from 0.35 in 2005 to 0.30 in 2020) and dropped below the level of segregation in female-dominated sectors after 2015.

\section{Data and Descriptive Statistics}\label{sec:data}
\subsection{Data Sources and Characteristics of the Sample}\label{subsec:data}
Our analysis is based on the Labour Force Survey (LFS) quarterly data from the UK Office for National Statistics (ONS). LFS is the most extensive household study in the UK, providing a comprehensive source of data on workers and the labour market. 
The analysis spans from fiscal years between 2005 and 2020. This period covers widespread enforcement of equality legislation, the 2007-2008 financial and economic crisis, and the recent changes caused by the \textsc{COVID}-19 outbreak.\footnote{Most of the literature shows that the 2007-2008 crisis had a severe impact on male-dominated sectors, such as on construction and manufacturing \citep{hoynes2012,Perivier2014,doepke2016}. In contrast, the COVID-19 crisis has hit counter-cyclical sectors (e.g., in-person services) sharply \citep{pilatowska2022}.}
 
We focus on working-age (16-64) employees, i.e., people who are in employment and paid a wage by an employer for their work.\footnote{We excluded self-employed workers (i.e., people who, in their main employment, work on their account whether or not they have employees) from the analysis because the earnings questions are not addressed to respondents who are self-employed in the UK LFS. This practice is standard in this type of analysis \citep[see, e.g.][]{dustmann2010,dustmann2014fiscal}.} Our final sample consists of 334,055 female workers and 307,245 male workers.
The dataset includes variables on a wide range of (i) demographic characteristics (gender, age, country of birth, nationality, ethnicity, religion); (ii) socio-economic factors (presence of dependent children, marital status, education, experience, full/part-time job, remote work, public sector, training opportunities, sectors and occupations); (iii) geographical information on residence and working region. 
Information on wages in the LFS is the self-reported gross weekly pay for the reference week.\footnote{We calculate the real wage based on hourly wages in 2015 prices as real wage = hour pay/(CPI2015/100).}
The number of hours is the total usual hours worked in the main job per week,  including usual hours of paid overtime when applicable. 


Table~\ref{tab:sum_fml} reports the summary statistics of the main variables by gender. 
Regarding demographic characteristics, there is a substantial prevalence of UK natives in both male and female samples (87\%), followed by non-European Economic Area (non-EEA) immigrants and EEA citizens. The average age is similar for men and women (around 40 years).  On average, women in the sample are as educated as men (13 years of education) and have slightly less experience (22.99 years of experience against 23.46). More than half of the women and men are either married or cohabiting. In addition, 46\% of women and 42\% of men have dependent children under 18 years.

Segregation variables show that 69\% of women are employed in female-dominated sectors compared to 35\% of men. Around 26\% of both women and men work in low-segregated sectors, while the rest work in highly segregated sectors.

Focusing on the outcomes of interest, both female and male employees have the same share of permanent jobs (around 94\%). Women's wages (in logarithm) are, on average, lower than men's (2.41 percentage points against 2.59 for men). 
A small share of employees work from home (5\% of the women against 8\% of the men). On average, women work 31 hours per week and  42\%  part-time, while men work  40 hours per week and 10\%  part-time.  
A more detailed investigation of the reasons for working part-time in Table~\ref{tab:sum_pt} highlights that the share of women who \emph{did not want} a full-time job is much larger than those who \emph{could not find} it  (78\% against 10\%). In contrast, 39\% of men  \emph{did not want} a full-time job and 26\%  \emph{could not find} one.
Family duties and domestic commitments are among the main reasons for not wanting a full-time job (48\% and 28\%, respectively). Instead, men do not want a full-time job mainly because \emph{they are financially secure and work because they want} (24\%) or for other reasons (34\%).



\subsection{Shift--Share Decomposition of Employment}\label{sec:ssdeco}

This section provides a descriptive picture of gender trends in employment in male-and-female-dominated sectors. We adopt a revised version of \citeapos{olivetti2016} shift-share decomposition.\footnote{Unlike the original paper that uses the number of worked hours, we use the employment shares.  \citet{razzu2020} present an extension of \citeapos{olivetti2016} decomposition considering the role of changing types of employment within industry sectors according to education from 1971 to 2016 in the UK.} 

In our analysis, the growth of female employment share is decomposed into (i) the change in the total \emph{employment share} of the sector (\emph{between component}), and (ii) the change in \emph{gender composition} within the sector (\emph{within component}):
\begin{equation}\label{eqn:deco_olivetti}
\Delta e_{st}^g= \underbrace{\sum_{j=1}^{J_s} \alpha_{jt}^g\Delta e_{jt}}_{\text{Between-sector}} + \underbrace{\sum_{j=1}^{J_s} \alpha_{jt} \Delta e_{jt}^g}_{\text{Within-sector}} \,\, \text{ for } s=\{fml-dom, ml-dom\}, g=\{fml, ml\}
\end{equation}

\noindent where $\Delta e_{st}^g = \frac{E_{st}^g}{E_{st}} - \frac{E_{t_0}^g}{E_{t_0}}$ is the difference in the share of female/male employment between the base period $t_0$ and the current period $t$; $\Delta e_{jt}= \frac{E_{jt}}{E_{t}} - \frac{E_{jt_0}}{E_{t_0}}$ is the difference in the share of total employment in sector $j$ between $t_0$ and $t$; $ \Delta e_{jt}^g= \frac{E_{jt}^g}{E_{jt}} - \frac{E_{jt_0}^g}{E_{jt_0}}$ is the difference in the share of female/male employment in sector $j$; $\alpha_{jt}^g = \frac{(e_{jt_0}^g+e_{jt}^g)}{2}$ and $\alpha_{jt} = \frac{(e_{jt_0}+e_{jt})}{2}$ are decomposition weights (i.e., the average share of female employment in sector $j$ and the average share of sector $j$, respectively). The reference year is the first available year in the dataset ($t_0=2005$); $s$ stands for sectors classified as female/male-dominated according to Equation~\eqref{eqn:frac}. 

\begin{figure}[ht!]
    \caption{Shift-share decomposition, by gender}\label{fig:shift_share}
    \centering
    \subfloat{\includegraphics[scale=.53]{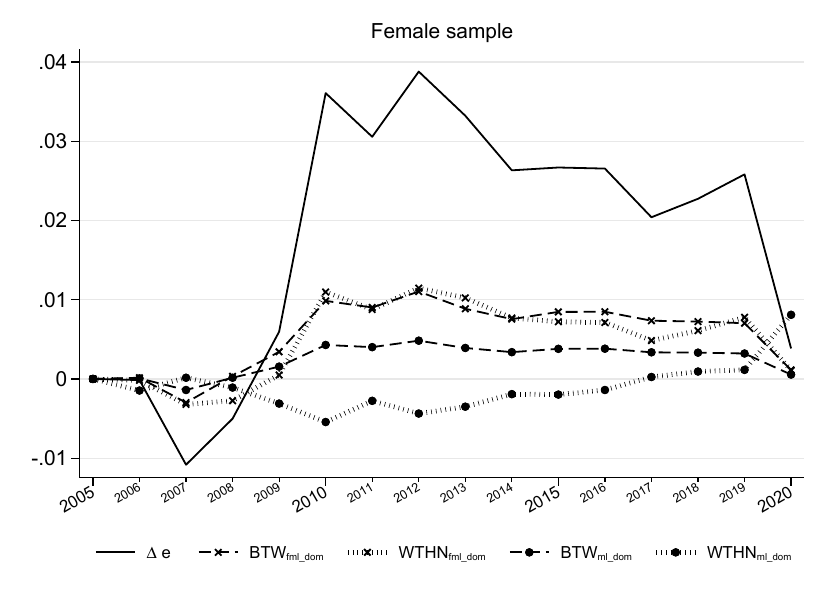}}\qquad
    \subfloat{\includegraphics[scale=.53]{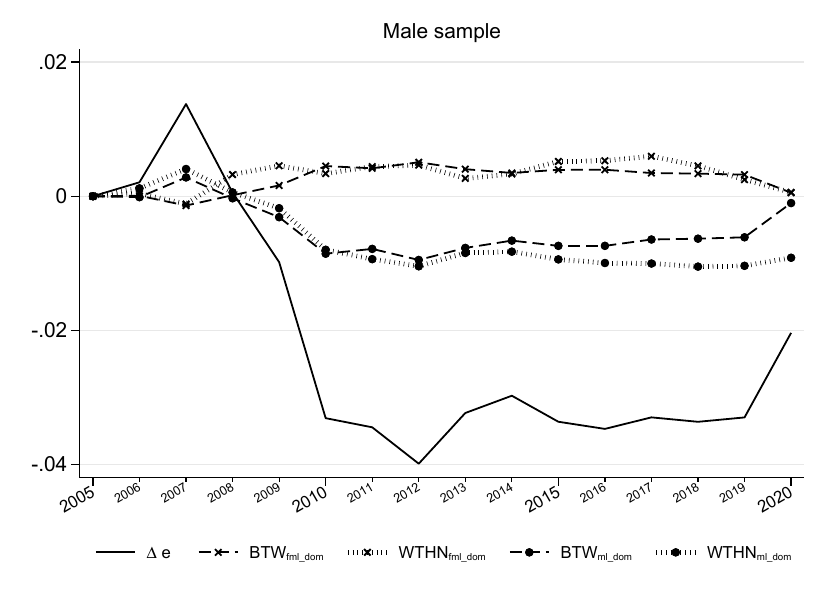}}\\
    \vspace{-1mm}
    \begin{minipage}{.95\linewidth}
    \vspace{-1mm}
    \footnotesize
    \emph{Note:  The graph shows the difference in employment in the comparison year with respect to the base year (i.e., the fiscal year 2005). The overall change in employment is shown in solid line and  its decomposition into the `between' and `within' components respectively, with dashed and dotted lines. The cross marks the components for  female-dominated sectors and the circle  the components  for male sectors. The `between' component (BTW) captures the change due to changes in the sectoral structure of the economy;  the `within' component (WTHN) reflects changes in female composition within sectors.}
    \end{minipage}
\end{figure}

Figure~\ref{fig:shift_share} displays the shift-share decomposition of female and male employment (left and right graphs, respectively). The graphs show the difference in employment with respect to the base year (2005) and its decomposition into  \emph{between}  and \emph{within} components (respectively, dashed and dotted lines) in female- and male-dominated sectors (crosses and circles). 

The 2007-2008 financial and economic crisis harshly hit male employment. In particular, male employment share (\emph{between} component) and male composition (\emph{within} component) in male-dominated sectors suddenly decreased. Conversely, the crisis stimulated female employment, not only in the total employment share  (both \emph{between} components of female- and male-dominated sectors) but also in the composition of women in female-dominated sectors (\emph{within} component). The \textsc{Covid--19} outbreak arrested the overall female employment in both male and female-dominated sectors (\emph{between} components). As expected,  this led to a reduction in the composition of women in female-dominated sectors (\emph{within} component) but to a sharp increase in the composition of women in male-dominated sectors.\footnote{These results are in line with the literature assessing that during the recession period of the 2007-2008 crisis, female employment was generally affected less than male employees, while during the recovery phase, male employment recovered faster than female employment \citep{hoynes2012,doepke2016,ellieroth2019,albanesi2021}. For instance, \citet{ellieroth2019} finds that married women are more stuck in employment during recessions. Therefore, their labour supply decisions account for the higher risk of job loss experienced by their husband. However, \citet{razzu2020} emphasises how gender segregation across industrial sectors and occupations in the UK exacerbates women's employment and pay gap during the business cycle.} 

Overall, female composition (\emph{within} component) increased gradually in male-dominated sectors after 2010. It seems that EA2010 stimulated female employment from the demand side. This may suggest that a higher proportion of women were employed \emph{within} each male-dominated sector at the expense of decreasing male employment. 

\section{Empirical Strategy}\label{sec:methods}
\subsection{\normalsize Estimating Gender Sectoral Segregation on Employment Contracts and Wages}\label{sec:employcontracts}

The aim of this section is to compare the average differences in labour market outcomes (permanent jobs, part-time jobs, working hours, remote work, and hourly wages) in male- and female-dominated sectors among workers with similar observable skills and socio-demographic characteristics. We implement the Propensity Score Matching (PSM) method, where ``working in female-dominated sectors'' is the assigned treatment. The underlying assumption is that those who choose to work in female- and male-dominated sectors only differ in the endowment of their observed skills and human capital accumulation. 

Let $p(\X_i)=Pr(D|\X_i)$ be the propensity score such that $p(\X_i)\in(0,1)$, where $D$ is the treatment and $\X_i$ a set of observable controls. The propensity scores obtained from a Probit regression model are used to match control and treated units.  
Provided that the assumptions of the PSM are satisfied, the formula of the average treatment effect on the treated (ATT) is
\begin{equation}
\begin{split}
\tau^{ATT} & \equiv \E\{Y_{1i}-Y_{0i}|D=1\} \\
& \E\Big[\E\{Y_{1i}-Y_{0i}|D=1, p(\X_i)\}\Big]\\
& \E\Big[\E\{Y_{1i}|D=1, p(\X_i)\} - \E\{Y_{0i}|D=0, p(\X_i)\}|D=1\Big]\\
\end{split}
\end{equation}
\noindent where $Y_{1i}$ and $Y_{0i}$ are the potential outcomes with and without treatment \citep[see][]{becker2002}. The comparison  \emph{between} gender-sectoral dominance is conducted for the full sample (male and female workers together), male sample, and female sample. 

For the best selection of covariates to calculate the propensity scores, we consider all factors associated with the treatment (including interactions and squares) and then an automatic selection procedure -- i.e., Least Absolute Shrinkage and Selection Operator (LASSO). Table~\ref{tab:lasso} in the Appendix reports the selected covariates from the penalised regressions. Then, we conduct the sensitivity analysis to test whether the balancing property of the covariates before and after the match holds. The covariates are balanced if the standardised bias after matching is within $\pm 5\%$ \citep{rosenbaum1985}. The matching method successfully builds a meaningful control group if the condition is satisfied. 

\subsection{Estimating Wages in Gender-Specific Dominated Sectors}
We now focus on the gendered differences in hourly wages in male- and female-dominated sectors based on observable and unobservable characteristics. For this purpose, we first perform the counterfactual Kitagawa-Blinder-Oaxaca (KBO) decomposition \citep{kitagawa1955,blinder1973,oaxaca1973}, which is an established method in the literature on discrimination \citep[e.g.,][]{mueller2006,blaukhan2017} to study the difference in wages between women and men. We then run Mincer wage regression to explore the role of the average characteristics of male and female employees in female- and male-dominated sectors and to obtain the predicted and residual wages.

\subsubsection{Decomposing the Gender Wage Differentials}

 In this analysis, we use the three-fold version of the KBO decomposition \citep{jann2008}, which decomposes the average difference of hourly wages (in logarithm) between men and women working in female- and male-dominated sectors into three components as follows:\footnote{We also implemented a standard two-fold KBO decomposition, which decomposes the wage differentials into a part that is explained by differences in human capital and productivity and an unexplained part usually interpreted as discrimination  \citep{blaukhan2017}. We obtained the same results for the explained and the unexplained components. However, because the unexplained component is the algebraic sum of the coefficient and the interaction effects of the three-fold decomposition, the two-fold is less informative \citep[see also][]{meara2020}.}
\begin{equation}\label{eqn:kbo}
\begin{split}
 \underbrace{\E(y^s_{ml}) -\E(y^s_{fml})}_{\text{overall difference}}  & = \underbrace{[\E(\X^s_{ml})-\E(\X^s_{fml})]' \bbeta^s_{fml}}_{\text{endowment effect}}  \\
& + \underbrace{\E(\X^s_{fml})'(\bbeta^s_{ml} -\bbeta^s_{fml})}_{\text{coefficients effect}} \\
& + \underbrace{[\E(\X^s_{ml})-\E(\X^s_{fml})]'(\bbeta^s_{ml} -\bbeta^s_{fml})}_{\text{interaction effect}} 
\end{split}
\end{equation}
\noindent where  $\X$ is a  vector containing the covariates, such as socio-demographic variables, human-capital variables, work-related variables; and $\bbeta$ is a vector of slope parameters and the intercept;  $fml$ stands for women and $ml$ for men, and \mbox{$s=\{fml-dom, ml-dom\}$}. 

The \emph{first component} explains observable group differences in the predictors, such as background and human capital characteristics of workers (\emph{endowment effect}). This effect quantifies the expected change in women's wages if they had men's characteristics. A negative \emph{endowment effect} shows that female workers possess better predictors than their male counterparts.

The \emph{second term} explains differences in the coefficients, including the intercept, that arise from discrimination -- i.e. unequal pay for equally qualified workers \citep{blaukhan2017} -- or cannot be explained by differences in the observed factors (\emph{coefficient effect}). Specifically, the \emph{coefficient effect} measures the expected change in the average wage of women if they had the coefficients of men. The intercept included in the effect captures the contribution of unobservable characteristics \citep{cotton1988} -- e.g., behavioural traits,  such as self-esteem, ambition, competitiveness and the willingness to take risky career choices \citep{gneezy2003, gneezy2004,Bertrand2011, saccardo2018}. A negative intercept term is interpreted as ``ongoing discriminatory constraints'', such as barriers in the labour market for the minority group due to the effects of discrimination and unobserved differences in productivity and tastes \citep{altonji1999}. When the overall \emph{coefficient effect} is positive, women would have higher average wages if paid like men.

The \emph{third component} explains the co-existence of differences in the endowments and coefficients between the two groups (\emph{interaction effect}).  If the \emph{interaction effect} is positive, women have a ``double disadvantage'' because they have smaller coefficients than men when they have worse predictors; if it is negative, differences in coefficients and covariate levels offset each other \citep{jann2018}.

After assessing which effect drives the wage differences,  in the next paragraphs, we investigate the contribution of each observable factor and unobservable characteristics that contribute to explaining the differences in wages between women and men in female- and male-dominated sectors.

\subsubsection{The Contribution of Human Capital}\label{sec:mincer}
We use the Mincerian regression to analyse the association of workers's human capital and observable skills with wage differentials between sectors and genders. The estimating equation is as follows:
\begin{equation}\label{eqn:mincer}
\mathbf{y} =  \mathbf{X}\bm{\beta} +  \delta_t +\bm{\epsilon}
\end{equation}

\noindent where $\mathbf{y} $ is hourly wages in logarithm; $\mathbf{X}$ is $N\times k$ matrix of control variables (i.e., socio-demographic, human-capital and work-related variables);  and $\delta_t$ are the time fixed effects. Equation~\eqref{eqn:mincer} is estimated using OLS. 

The set of controls includes three groups of variables as follows. \emph{Socio-demographic variables} include age and its square, nationality, ethnicity, religion, being in a stable relationship, having dependent children and the interaction of the last two. \emph{Human-capital variables} are education (low, intermediate, and higher education), experience and its square, years in education and its square, and training offered by the current employer.\footnote{We included both the categorical variable for the education band (low, intermediate and high) and the continuous variable for years of education. The OLS assumption of the absence of perfect multicollinearity is not violated because years of education capture the intensity of the returns of education within each band.}  \emph{Work-related variables} include a dummy for female-dominated sectors, a dummy for low gender sector segregation, a dummy for working in the public sector, and the type of occupation. Working region dummies are included.\footnote{Usual worked hours are not included in the specification because of possible endogeneity issues due to reverse causality. The estimates would be downward biased because hours would appear on both sides of the equation as wages are calculated based on usual worked hours per week \citep{borjas1980}.}

Mincer regression is based on observable characteristics so as to ``hold constant'' individual factors that affect wages. However, this specification may not capture the selection of workers into sectors based on some relevant unobservable characteristics (e.g. self-esteem, ambition, competitiveness, risk aversion) that may influence wage differentials. In the next section, we address the possible bias that arises from omitting these unobservable factors. 


\subsubsection{The Role of Predicted and Residual Wages}\label{sec:unobservable}
We use the estimates of Mincer regression from Section~\ref{sec:mincer} to measure how the selection on observable and unobservable characteristics shapes the difference in wages for men and women in female- and male-dominated sectors. First, we calculate the estimated returns to construct predicted wages, which measure the individual wage potential based on observable factors \citep{parey2017}. Second, we follow \citet{borjas2019} to shed light on the role of unobservable characteristics in the selection process by calculating residual wages, which capture the part of wages uncorrelated with workers' skills. 

We consider four sub-groups from our sample of workers: men in male-dominated sectors (\emph{ml}, \emph{ml-dom}), women in male-dominated sectors (\emph{fml}, \emph{ml-dom}), men in female-dominated sectors (\emph{ml}, \emph{fml-dom}), women in female-dominated sectors (\emph{fml}, \emph{fml-dom}). 
In addition, we conduct a counterfactual exercise to examine the trajectory of wage potentials and residuals for each sub-group if workers had the same estimated coefficients of men working in male-dominated sectors. In formulae,
\begin{align}
&\hat{\mathbf{y}}^c_{g, gdom}= \X_{g, gdom}\bm{\hat{\beta}}_{g, gdom}\\
&\hat{\mathbf{u}}^c_{g, gdom}= \mathbf{y}_{g, gdom}-\hat{\mathbf{y}}^c_{g, gdom}
\end{align}
\noindent where $g=\{ml, fml\}$ is the gender of the worker, and $gdom=\{ml-dom, fml-dom\}$ is the sector with a large share of the specified gender. 
This analysis allows us to compare how predicted and residual wages would differ if workers (women in female- and male-dominated sectors and men in female-dominated sectors) had the same characteristics of the most advantaged group, i.e. men in male-dominated sectors.

Predicted and residual wages are sorted and used to construct the Cumulative Distribution Functions (CDFs) by gender in female- and male-dominated sectors. Then, we compare the CDFs of men and women \emph{between} and \emph{within} gender-sectoral dominance.  
The equality of the distributions of the (actual and counterfactual) predicted and residual wages among the four sub-groups is tested by using the non-parametric Kolmogorov-Smirnov (K-S) test. 

\section{Estimation Results}\label{sec:results}
\subsection{Estimation Results for the PSM on Contracts and Wages}\label{sec:res.contracts}
In this section, we present the main results of the PSM by looking at three different samples (i.e.,  all workers, men, and women). Working in a female-dominated sector is the treatment variable. Table~\ref{tab:att} reports the ATT  for each labour market outcome of interest -- i.e., having a temporary job, part-time work, number of hours (in logarithm), remote work, and wage (in logarithm).\footnote{%
The propensity scores for matching treated and control units come from estimates reported in Table~\ref{tab:probit_fmldom} in the Appendix. The table shows the likelihood of a worker being employed in a female-dominated sector based on socio-demographic characteristics and working environment. Being a woman in a stable relationship without dependent children decreases the probability of working in a female-dominated sector. Having dependent children, being non-European and working in operative jobs, technical and secretarial occupations reduces the likelihood of being in female-dominated sectors.} 

The first result of the analysis is that contractual features usually associated with female workers are more common in female-dominated sectors, even among men. That is, workers in female-dominated sectors compared to their peers in male-dominated sectors have, on average, fewer permanent positions (respectively, 0.947 against 0.956), work more part-time (0.35 against 0.267), fewer hours (3.404 against 3.475), and less from home (0.037 against 0.90). This result remains valid even when looking at male and female workers separately.  Specifically, if men and women in a female-dominated sector were hired in a male-dominated sector, they would have more permanent positions (0.9 p.p. and 0.7 p.p., respectively),  would work more hours (7.7 and 6.3 p.p.), less part-time (8.4 and 7.5 p.p.), and more from home (4.3 and 5.9 p.p.). All ATT estimates are significantly different from zero at a 1\% significance level.

The second main result is that there is a higher penalty for men than women working in female-dominated sectors, given their larger magnitudes of ATT (all significant at 1\% level). This result is also confirmed when looking at wage differentials between female- and male-dominated sectors. In fact, men in female-dominated sectors earn, on average, 15.4 p.p. less than their male peers in male-dominated sectors. Instead, women in female-dominated sectors earn, on average, 12.6 p.p. less than their female counterparts in male-dominated sectors. Overall, any worker in female-dominated sectors would be paid 13.6 p.p. more if employed in male-dominated sectors.
These results are consistent with the findings of ``comparable worth'' literature, that is, jobs dominated by women pay, on average, less all employees  \citep{treiman1981, killingsworth1987}, and the effect on wages in such jobs is more negative for men than women \citep{roos1981, johnson1984}.

The sensitivity analysis (Figure~\ref{fig:balance} in the Appendix) confirms that the balancing property is satisfied for all samples since all covariates are well balanced.  Therefore, the matching method effectively built a valid control group. 

Overall, our analysis suggests that gender sectoral segregation is a relevant factor in explaining observed differences in employment contracts (i.e. part-time, permanent, remote work, number of weekly working hours) and wage differentials.

\subsection{Estimation Results for Wages}
\subsubsection{Results for the KBO}
\begin{figure}[th!] 
\vspace{8mm}
	 \caption{KBO decomposition, by gender sectoral dominance}\label{fig:oaxaca_dom}
  \centering
	\includegraphics[scale=1]{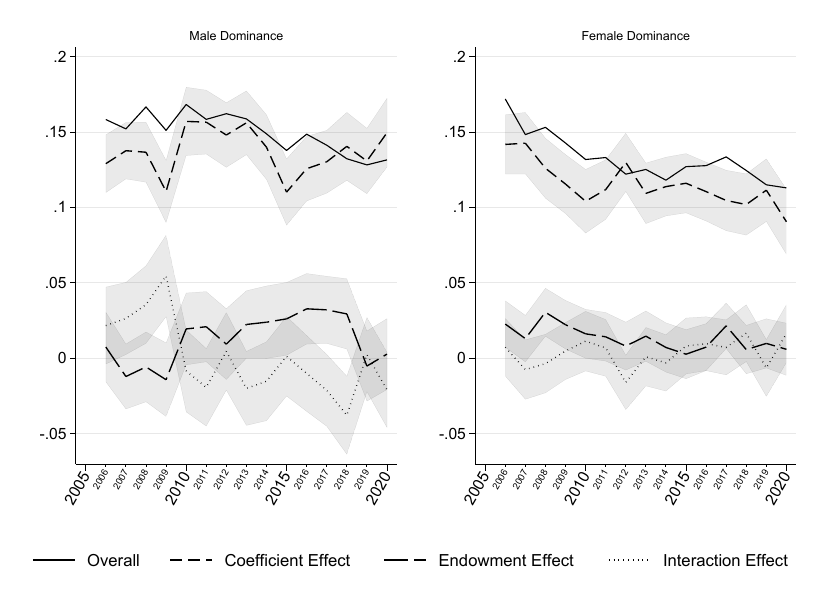}\\
	 \begin{minipage}{.95\linewidth}
	\vspace{3mm}
	\footnotesize
	\emph{\textsc{Estimation note}: Both models for women and men are estimated using the Mincerian regression equation (with OLS). The degree of gender segregation is not included because it is highly correlated with the grouping variable of gender sectoral dominance. The shaded areas are the 95\% confidence intervals.}
	\end{minipage}
\end{figure}

The evolution of the three components of the KBO decomposition and their sum over time is shown in Figure~\ref{fig:oaxaca_dom}. Women are contrasted to men within the same gender-dominated sector. The dashed line represents the \emph{coefficient effect}, the long-dashed line the \emph{endowment effect} and the dotted line the part of the  \emph{interaction} component. The solid line is the sum of the three effects and reveals their overall contribution.\footnote{For the contribution of each characteristics, see Tables \ref{tab:oaxaca_factors_fmldom}-\ref{tab:oaxaca_factors_mldom} in the Appendix.} 

The first result from the decomposition is that the difference in wages between men and women in both types of sectors is not so much explained by differences in human capital and productivity (\emph{endowment effect}). The dynamics of the \emph{endowment effect} shows that the gap in terms of observable characteristics has narrowed over time (the effect is close to zero), reflecting women's increased human-capital levels relative to men's  \citep[as also observed by, e.g.,][]{goldin2014,blaukhan2017}. While men and women employed in female-dominated sectors are, on average, more similar in terms of human capital over time, the \emph{endowment effect} is positive between 2010 and 2018 in male-dominated sectors, meaning that women have worse observed characteristics than men in those years.

Instead, the most relevant result is that the difference is mostly due to ``ongoing discriminatory constraints'' in the labour market towards women (\emph{coefficient effect}) stemming from substantial unexplained constraints in labour market returns (the intercept is negative from Tables \ref{tab:oaxaca_factors_fmldom}-\ref{tab:oaxaca_factors_mldom} in the Appendix). The \emph{coefficient effect} is positive in both gender-dominated sectors, suggesting that women should be paid more than men to prevent any sort of discrimination for reasons other than human capital characteristics and productivity. 

The  \emph{interaction effect} explains little of the gender wage differential in both female- and male-dominated sectors, although we observe a ``double disadvantage''  for women (positive \emph{interaction effect})  in male-dominated sectors only before 2010.



\subsubsection{Results based on Human Capital Factors}
As the KBO showed that human capital characteristics play a minor role in explaining wage differentials,  the analysis in this section shows the contribution of each observable factor on wages. Table~\ref{tab:mincer_regs} reports the estimated coefficients of the Mincer wage regression by gender for all sectors (Columns 1-2) and gender-dominated sectors (Columns 3-6).

The estimates of the Mincer regression for segregation variables confirm the main result of the PSM. Specifically,  working in female-dominated sectors is significantly negatively correlated with hourly wages for both men and women (-0.163 and -0.158, respectively).  In addition, working in sectors with low gender sectoral segregation is significantly positively associated with higher wages for male workers in the full sample (0.027) but negatively correlated with wages for women in both female- and male-dominated sectors (-0.012 and -0.102, respectively). The interaction term between female-dominated sectors and low gender segregation is positive and significant for women only. 

Focusing on human capital characteristics, workers with higher educational attainment earn, as expected, more than those with low education.  More years of education are positively associated with wages but with a diminishing effect (the square is negative). Our calculations show that the optimal number of years in education that maximises wages is approximately 15.7 years for men as opposed to 19.5 years for women in the full sample.\footnote{The figures come from the following calculations: $0.157/(2\times0.005)=15.7$ for men, and $ 0.117/(2\times0.003)=19.5$ for women.} Therefore, women are expected to stay in education for more years than men, who need only a degree to earn their optimal wage. We obtained a similar number of years of education in female-dominated sectors (18 for women and 15.9 years for men), while the difference is less pronounced in male-dominated sectors (16.5 for women and 15.6 for men). Potential working experience has significant diminishing returns, and receiving training is significantly associated with an increase in the hourly wage, especially in male-dominated sectors.

For socio-demographic and job characteristics, being non-UK natives is significantly associated with lower wages. However, the reduction is, on average, larger in absolute terms for EEA than non-EEA, except for female-dominated sectors. The presence of dependent children penalises women's wages but not men's, independently of the sector. Further, working in the public rather than in the private sector is associated with higher wages for women than men. However, the coefficients are non-significant in male-dominated sectors. This suggests that the private sector pays more in male-dominated sectors while the public sector offers better remuneration in female-dominated sectors. 




\subsubsection{Results based on Predicted and Residual Wages}\label{sec:res.unobservable}
\begin{figure}[tb!]
	 \caption{CDFs of predicted wages, by gender and sectoral dominance}\label{fig:cdf_yhat}
  \centering
	{\includegraphics[scale=1.05]{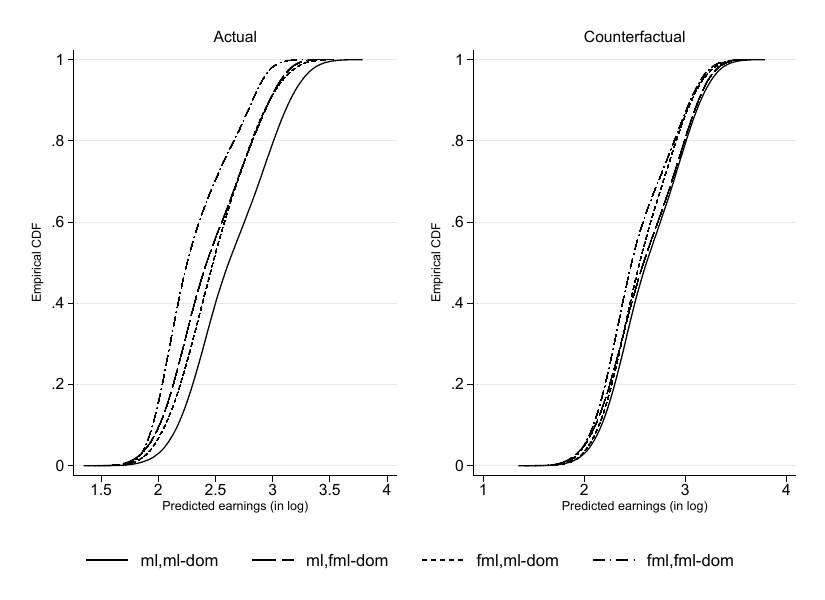}}\\
		 \begin{minipage}{.85\linewidth}
	\vspace{3mm}
	\footnotesize
	\emph{Note: The solid line is for men working in male-dominated sectors, the short-dashed line is for women employed in male-dominated sectors, the long-dash line is for men in female-dominated sectors, and the dash-dot line is for women in female-dominated sectors. Left: Predicted wages are calculated after estimating the coefficients of the Mincerian wage regression, reported in Table~\ref{tab:mincer_regs}. Right: predicted wages are calculates using  the estimated coefficients from the  Mincerian regression of men working in male-dominated sectors. Predicted wages in the counterfactual exercise are precise measure of individual earnings potential  \citep{gould2016,borjas2019}.}
	\end{minipage}
\end{figure}

\begin{figure}[tb!]
	 \caption{CDFs of residual wages, by gender and sectoral dominance}\label{fig:cdf_uhat}
  \centering
	{\includegraphics[scale=1.05]{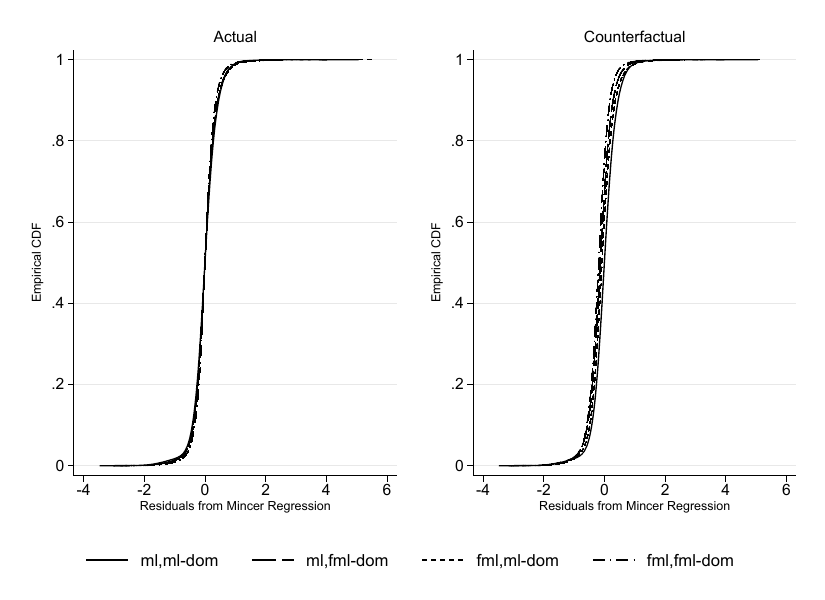}}\\
		 \begin{minipage}{.85\linewidth}
	\vspace{3mm}
	\footnotesize
	\emph{Note: The solid line is for men working in male-dominated sectors, the short-dashed line is for women employed in male-dominated sectors, the long-dash line is for men in female-dominated sectors, and the dash-dot line is for women in female-dominated sectors. Left: Residual wages are calculated after estimating the coefficients of the Mincerian wage regression, reported in Table~\ref{tab:mincer_regs}. Right: Residual wages are calculates using  the estimated coefficients from the  Mincerian regression of men working in male-dominated sectors. Residuals from a Mincerian regression calculated in this way capture the part of earnings that is uncorrelated to observed skills \citep{parey2017}.}
	\end{minipage}
\end{figure}

This section discusses empirical evidence on the differences in the selection of workers in male- and female-dominated sectors in terms of observable (predicted wages) and unobservable (residual wages) characteristics. 
Figures~\ref{fig:cdf_yhat} and~\ref{fig:cdf_uhat} respectively display the CDFs of potential and residual wages for the four subgroups: men in male-dominated sectors (\emph{ml}, \emph{ml-dom}), women in male-dominated sectors (\emph{fml}, \emph{ml-dom}), men in female-dominated sectors (\emph{ml}, \emph{fml-dom}), women in female-dominated sectors (\emph{fml}, \emph{fml-dom}). The graphs on the left show actual values, calculated using the estimated coefficients for each subgroup from Table~\ref{tab:mincer_regs}. The graphs on the right display counterfactual values calculated with the estimated coefficients of men working in male-dominated sectors.



The key result from the left graph in Figure~\ref{fig:cdf_yhat} is that there is a penalty in potential wages associated with working in a female-dominated sector. In fact, women in female-dominated sectors always have lower predicted wages than all other workers (CDFs always lying on the left). For low levels of potential wages,  men employed in female-dominated sectors earn much less than women in male-dominated sectors. 

If workers had the potential wages of men in male-dominated sectors,  wage differentials of men and women across female- and male-dominated sectors would be smaller (Figure~\ref{fig:cdf_yhat} on the right). However, women in female-dominated sectors would always be paid less than all other workers. Men in female-dominated sectors would still be penalised compared to workers in male-dominated sectors, but only in low-paid jobs. But moving to the top of the distribution, the gap in terms of potential counterfactual wages between women and men increases, meaning that women would always earn less than men. These findings contrast  \citet{roos1981} and \citet{johnson1984}, who always find a more pronounced wage differential for men than women in female-dominated environments.

When we look at the residual wages in Figure~\ref{fig:cdf_uhat}, the results highlight that differences in wages in high-paid jobs cannot be attributed to acquired skills or accumulated human capital only. Specifically, the CDF of women in female-dominated sectors (left graph) lies to the right of the other curves for low residual wages (positive selection at the bottom of the distribution) and to their left for high values (negative selection at the top). This means that these women earn more in low-paid jobs but much less in high-paid jobs than the other workers for reasons other than their skills and human capital. 

In the counterfactual exercise (right graph), all curves would shift to the left of the CDF of male workers in male-dominated sectors, meaning that all workers would be negatively selected with respect to the former. At the bottom of the distribution, both women and men in female-dominated sectors would be penalised in terms of residual wages due to unobserved characteristics (the two CDFs overlap and lie to the left of the other two). However, as we move up to the distribution, the CDFs diverge, and women in female-dominated sectors are more negatively selected (laying more to the left) than their male counterparts and other subgroups.

The non-parametric K-S test always rejects the null hypothesis of equality of distributions among the four sub-groups (see Table~\ref{tab:ksmirnov} in the Appendix), confirming that the distributions of  (actual and counterfactual) predicted and residual wages of men and women across sectors differ.



\section{Conclusion}\label{sec:conclusion}

This paper studied the contribution of gender sectoral segregation in explaining the differences in contracts (i.e., permanent jobs, part-time jobs, working hours, remote work) and hourly wages in the UK between  2005 and 2020. We further analysed how wages differ in female- and male-dominated sectors by looking at both observable and unobservable characteristics.


Our empirical analysis suggests that the persistent imbalance in the shares of men or women in some sectors contributes to explaining the differences in employment contracts and wages.

We first found that female-dominated sectors reflect contractual characteristics typical of women. In other words, working in female-dominated sectors is associated with a greater reliance on part-time contracts, fewer hours, and less working from home, even controlling for the occupational composition. In addition, female-dominated sectors seem to pay, in general, less for any worker regardless of their gender. The penalty for men working in female-dominated sectors is even larger than for women.

Second, women working in female- and male-dominated sectors are paid less not because of differences in human capital and productivity but rather because of the existence of persistent ``discriminatory constraints'', such as barriers in the labour market for women due to the effects of discrimination and unobserved differences in productivity and tastes. This means that women have observable attributes similar to men regarding accumulated human capital, and without these discriminatory barriers, wage differentials between women and men within male- and female-dominated sectors would be lower. 

Third, female-dominated sectors are not as rewarding as male-dominated sectors in terms of predicted and residual wages. While women in female-dominated sectors are always worse off than all other workers, men in female-dominated sectors are disadvantaged in low-paid jobs only. In addition, actual and counterfactual results for residual wages have documented the negative selection of women in female-dominated sectors with respect to all other workers, especially at the top of the wage distribution. The use of predicted and residual wages allows us to control the issue of selection based on unobservables that arises from the use of the PSM and Mincer regression, where the former matches workers with similar observed characteristics and the latter holds constant observed individual factors associated with wages.

This analysis has policy implications.  Gender segregation in the labour market may be responsible for causing more challenges for women than their male counterparts regarding labour participation, access to jobs and career opportunities. This gap could potentially widen in the post-pandemic.  Our findings can provide policy-makers with empirical evidence supporting appropriate reforms favouring vulnerable categories of workers (i.e., women, mothers, and immigrants) and policies designed to sustain long-run economic growth, especially as the UK is facing new challenges (i.e., pandemic and Brexit).  Future avenues of research could focus on gender segregation into sectors that have seen a rise in the use of atypical work arrangements, e.g., zero-hour contracts and casual work.

\clearpage
\begin{spacing}{1}
\bibliographystyle{apalike}
\bibliography{ch3_gender_LMP.bib}
\end{spacing}
\clearpage
\section{Tables in Text}

\begin{table}[th]
\centering
\caption{Share of women, by sector and year}\label{tab:sectors1}
\scalebox{.8}{
\begin{threeparttable} 	
   \begin{tabular}{lccccc}
\vspace{-3mm}\\   
\hline\hline
\vspace{-3mm}\\
\multicolumn{1}{c}{Sectors}&\multicolumn{5}{c}{Women's Share (\%)}\\
\cmidrule(l{.25cm}r{.25cm}){2-6}
& 2005&2010&2015&2020&2005-2020\\
\hline
\vspace{-2mm}\\	
A - Agriculture, forestry \& fishing&30.8&25.4&32.7&31.9&29.9\\
B - Mining \& quarrying&15.3&13.8&15.2&23.4&18.0\\
C - Manufacturing&25.6&24.6&25.7&28.8&25.9\\
D - Electricity, gas \& air con supply&24.6&25.1&27.1&26.7&27.6\\
E - Water supply, sewerage \& waste&21.3&18.8&21.7&23.2&20.3\\
F - Construction&14.6&16.8&18.4&21.1&16.8\\
G - Distribution &53.4&51.5&51.2&49.3&51.7 \\
H - Transport \& storage&26.4&23.0&25.1&24.8&24.5\\
I - Accommodation \& food services &58.6&57.9&56.0&57.9&57.5 \\
J - Information \& communication&27.8&30.9&29.5&32.6&30.4\\
K - Financial \& insurance services&54.2&51.0&50.1&48.3&50.9\\
L - Real estate services &56.2&63.2&55.3&58.3&57.8 \\
M - Professional, scientific \& technical activities&49.8&47.7&48.2&46.8&48.0\\
N - Admin \& support services&24.5&46.9&49.7&48.8&44.6\\
P - Education &74.3&75.6&74.8&76.1&75.3\\
Q - Health \& social work &80.5&80.6&80.3&79.1&80.4\\
R - Arts, entertainment \& recreation&50.0&52.4&51.3&50.7&50.3\\
S - Other service activities &66.1&61.7&61.8&60.6&62.4 \\
T - Households as employers &68.6&78.4&79.6&77.1&74.8 \\
\hline
	\end{tabular}	
\begin{tablenotes}[para,flushleft]	
\emph{Notes: Sectors labelled as ``O - Public admin \& defense'' and ``U - Extra territorial'' are removed from the sample because their contracts and wages highly differ from other sectors. The share is calculated over total employment whereas gender sectoral dominance is calculated with the Sectoral Segregation Index in Equation~\eqref{eqn:ssi}.}
  \end{tablenotes}
  \end{threeparttable}
  }
\end{table}
\begin{table}[th]
\centering
\caption{List of high and low segregated sectors}\label{tab:sectors2}
\scalebox{.8}{
\begin{threeparttable} 	
   \begin{tabular}{ll}
\vspace{-3mm}\\   
\hline\hline
\vspace{-3mm}\\

\multicolumn{1}{c}{High segregated sectors}&\multicolumn{1}{c}{Low segregated  sectors}\\
\hline
\vspace{-2mm}\\	
\multicolumn{2}{l}{\emph{Female-dominated sectors}}\\
I - Accommodation \& food services &  G - Distribution \\
P - Education& L - Real estate services \\
Q - Health \& social work &S - Other service activities \\
 &T - Households as employers \\

\vspace{-2mm}\\	
\multicolumn{2}{l}{\emph{Male-dominated sectors}}\\
 B - Mining \& quarrying& A - Agriculture, forestry \& fishing\\
C - Manufacturing & K - Financial \& insurance services \\
D - Electricity, gas \& air con supply&R - Arts, entertainment \& recreation\\
E - Water supply, sewerage \& waste &\\
F - Construction &\\
H - Transport \& Storage &  \\
J - Information \& communication  & \\
M - Professional, scientific \& technical activities&   \\
  N - Admin \& support services &\\
& \\

  \hline\hline
	\end{tabular}	
\begin{tablenotes}[para,flushleft]	
\emph{Notes: Sectors labelled as ``O - Public admin \& defense'' and ``U - Extra territorial'' are removed from the sample because their contracts and wages highly differ from other sectors.}
  \end{tablenotes}

  \end{threeparttable}
  }
\end{table}

 \begin{table}[th!]
\centering
\caption{Summary statistics for employed workers}\label{tab:sum_fml}
\scalebox{.8}{
\begin{threeparttable} 	
\begin{tabular}{lcccccc}
\hline\hline
\vspace{-3mm}\\
 &\multicolumn{2}{c}{Women}&\multicolumn{2}{c}{Men}&\multicolumn{2}{c}{Full sample} \\
\cmidrule(l{.35cm}r{.25cm}){2-3}\cmidrule(l{.35cm}r{.25cm}){4-5}\cmidrule(l{.35cm}r{.25cm}){6-7}
Variable & Mean& Std. dev.&   Mean& Std. dev.&   Mean& Std. dev.\\
\hline
\vspace{-3mm}\\
\multicolumn{6}{l}{\emph{Demographic characteristics}}\\
Natives&   0.87&        0.33&        0.87&        0.34&     0.87&        0.33\\
EEA&       0.05&        0.22&          0.05&        0.22&        0.05&        0.22\\
non-EEA&      0.07&        0.26&         0.08&        0.27&         0.08&        0.27\\
Age&  40.58&       11.99&          41.15&       12.37&   40.85&       12.18\\
Black&       0.02&        0.14&    0.02&        0.13&  0.02&        0.14\\
Asian&         0.03&        0.17&       0.04&        0.20&     0.04&        0.18\\
Other ethnicity  &        0.03&        0.16&     0.03&        0.16&      0.03&        0.16\\
Muslim&     0.01&        0.12&          0.03&        0.16&      0.02&        0.14\\
Christian&  0.61&        0.49&        0.54&        0.50& 0.57&        0.50\\
Other religions &    0.14&        0.35&     0.14&        0.34&    0.14&        0.35\\

\vspace{-3mm}\\
\multicolumn{6}{l}{\emph{Socio-economic factors}}\\
In couple    &   0.55&        0.50&    0.58&        0.49&        0.57&        0.50\\
With dependent children (<18yrs)&  0.46&        0.50&    0.42&        0.49&           0.44&        0.50\\
Years of Education&   13.77&        3.07&   13.58&        2.91&   13.68&        2.99\\
Experience  &    22.99&       12.44&   23.46&       12.83&   23.21&       12.63\\
Training     &     0.34&        0.47&    0.33&        0.47&     0.34&        0.47\\
Log wages   & 2.41&        0.50&         2.59&        0.56&      2.50&        0.54\\

Part-time work &       0.42&        0.49&    0.10&        0.29&     0.26&        0.44\\
Public sector     &  0.34&        0.47&     0.14&        0.35&  0.24&        0.43\\
Permanent job  & 0.94&        0.23&      0.95&        0.21&   0.95&        0.22\\

Weekly hours & 30.77&       12.87&   40.19&       12.53&    35.28&       13.55\\
Log weekly hours&3.32&        0.51&      3.64&        0.37&3.47&        0.48\\
Remote work&   0.05&        0.21&     0.08&        0.26&    0.06&        0.24\\
Benefit& 0.40&        0.49&      0.11&        0.31&   0.26&        0.44\\
Female dominance&  0.69&        0.46&     0.35&        0.48&         0.53&        0.50\\
Low segregation &   0.27&        0.44&0.26&        0.44&       0.26&        0.44\\

\hline
\end{tabular}
\begin{tablenotes}[para,flushleft]	
\emph{Notes: The descriptive statistics in the table are for the sample of employed workers (self-employed are excluded from the sample). Total number of female workers is 334,055.  Total number of male workers is 307,245. Total number of workers in the sample is 641,300.}
  \end{tablenotes}
  \end{threeparttable}
  }
\end{table}

 \begin{table}[th!]
\centering
\caption{Reasons for part-time work}\label{tab:sum_pt}
\scalebox{.85}{
\begin{threeparttable} 	
\begin{tabular}{lccc}
\hline\hline
\vspace{-3mm}\\
& Women (\%)&Men (\%)&Full sample (\%)\\
\hline
\vspace{-3mm}\\
\multicolumn{4}{c}{\emph{Reasons for part-time work}}\\
\vspace{-3mm}\\
Student or at school& 9.69& 30.57 &  13.33 \\
Ill or disabled&   2.15& 4.65 &2.59\\
Could not find full-time job&9.95 & 25.88 &12.74 \\
Did not want full-time job&  78.21   & 38.91&71.34\\
Total &100&100&100\\
\vspace{-3mm}\\
\hline
\vspace{-3mm}\\
\multicolumn{4}{c}{\emph{Reasons for not wanting full-time job}}\\
\vspace{-3mm}\\
Financially secure -- work because want & 7.22&24.26&8.74\\
Earn enough part-time& 7.64& 17.41&  8.52\\
Want to spend more time with family & 40.35  &12.30  &37.85 \\
Domestic commitments prevent full-time& 28.20&10.70 & 26.64 \\
Insufficient child-care facilities & 3.49& 0.96 &  3.26\\
Another reason& 13.09   &34.37&14.99 \\
Total &100&100&100\\
\hline
\end{tabular}
\begin{tablenotes}[para,flushleft]	
\emph{Notes: Percentages (\%) are over group total.}
  \end{tablenotes}
  \end{threeparttable}
  }
\end{table}

\begin{table}[tp!]
\centering
\caption{Propensity score matching}\label{tab:att}
\scalebox{.8}{
\begin{threeparttable} 	
   \begin{tabular}{lccccccc}
\vspace{-3mm}\\   
\hline\hline
\vspace{-3mm}\\
Variable &Treated &Controls &\shortstack{Difference \\ (ATT)}& S.E. &T-stat&\shortstack{Untreated units \\ on support}& \shortstack{Treated units \\on support} \\
  \hline
  \vspace{-3mm}\\
  \multicolumn{8}{l}{\emph{Sample: Full }}\\
Permanent&0.947&  0.956&-0.009&0.001& -10.44&295,963& 321,014\\ 
Part-time work&0.350&0.267&0.082&0.002&51.98&295,994&321,056\\
ln(hours)&3.404& 3.475&  -0.071&0.002&-43.64&296,025&321,094 \\
Remote work&0.037&  0.090& -0.053&0.001& -46.70&296,025&321,094\\
ln(wage)&2.419&2.555&-0.136&0.002&-59.81&296,025& 321,094 \\
  \vspace{-3mm}\\
  
  \multicolumn{8}{l}{\emph{Sample: Men}}\\
Permanent&0.949&0.958&-0.009&0.001&-9.51&  195,607&100,778\\
Part-time work&0.144&0.060&0.084&0.001&63.27&195,627&100,791\\
ln(hours) &3.597& 3.674&  -0.077&0.002&-50.52&  195,650&100,802 \\
Remote work&0.049&  0.092& -0.043&0.001& -33.90& 195,650 &100,802\\
ln(wage)&2.502&2.656&-0.154&0.003&-58.40& 195,650& 100,802 \\
 
    \vspace{-3mm}\\
  \multicolumn{8}{l}{\emph{Sample: Women}}\\
Permanent&0.947&0.954&-0.007&0.001&-6.44&100,356&220,214\\
Part-time work&0.444&0.369&0.075&0.003&28.27& 100,367&220,243\\
ln(hours) &3.315& 3.378&  -0.063&0.003&-24.51&100,375&220,270\\
Remote work&0.031&  0.090& -0.059&0.002& -39.18&100,375&220,270\\
ln(wage)&2.380&2.506&-0.126&0.003&-42.40&100,375&220,270\\
\vspace{-3mm}\\          
  \hline\hline
	\end{tabular}	
\begin{tablenotes}[para,flushleft]	
\emph{Note: S.E. does not take into account that the propensity score is estimated. The matching method is single nearest-neighbour; five neighbors are used to calculate the matched outcome. The matching algorithm imposes common support. The propensity scores obtained from Probit regression in Table~\ref{tab:probit_fmldom} are used to match control and treated units. The estimates are significantly different from zero at a 1\% significance level.}
  \end{tablenotes}
  \end{threeparttable}
  }
\end{table}

 \begin{table}[tp!]
\centering
\caption{Mincerian regression results, years 2005-2020}\label{tab:mincer_regs}
\scalebox{.75}{
\begin{threeparttable} 	
   \begin{tabular}{lcccccc}
\vspace{-3mm}\\   
\hline\hline
\vspace{-3mm}\\
&\multicolumn{6}{c}{\emph{Dep. var.:} Log(Wage)} \\
\cmidrule(l{.35cm}r{.25cm}){2-7}
\vspace{-3mm}\\
&\multicolumn{2}{c}{All sectors}&\multicolumn{2}{c}{Male-dominated sectors}&\multicolumn{2}{c}{Female-dominated sectors}\\
\cmidrule(l{.35cm}r{.25cm}){2-3}\cmidrule(l{.35cm}r{.25cm}){4-5}\cmidrule(l{.35cm}r{.25cm}){6-7}
\vspace{-4mm}\\
&Man &Women &Man &Women&Man &Women \\
&(1)&(2)&(3)&(4)&(5)&(6)\\

\hline 
\vspace{-2mm}\\
\multicolumn{7}{l}{\emph{Human capital variables}}\\            
Intermediate education &           0.022***&       0.020***&       0.022***&       0.025***&       0.019***&       0.017***\\
            &     (0.003)   &     (0.002)   &     (0.004)   &     (0.005)   &     (0.004)   &     (0.003)   \\
High education &          0.099***&       0.100***&       0.100***&       0.103***&       0.090***&       0.099***\\
            &     (0.006)   &     (0.005)   &     (0.007)   &     (0.010)   &     (0.009)   &     (0.006)   \\
Years of education   &      0.157***&       0.117***&       0.156***&       0.132***&       0.127***&       0.108***\\
            &     (0.003)   &     (0.003)   &     (0.004)   &     (0.006)   &     (0.006)   &     (0.003)   \\
Years of education$^2$&    -0.005***&      -0.003***&      -0.005***&      -0.004***&      -0.004***&      -0.003***\\
            &     (0.000)   &     (0.000)   &     (0.000)   &     (0.000)   &     (0.000)   &     (0.000)   \\
Experience  &      0.015***&       0.013***&       0.017***&       0.016***&       0.013***&       0.012***\\
            &     (0.001)   &     (0.001)   &     (0.001)   &     (0.002)   &     (0.001)   &     (0.001)   \\
Experience$^2$&       -0.000***&      -0.000***&      -0.000***&      -0.000***&      -0.000***&      -0.000***\\
            &     (0.000)   &     (0.000)   &     (0.000)   &     (0.000)   &     (0.000)   &     (0.000)   \\
Training &       0.067***&       0.049***&       0.068***&       0.063***&       0.047***&       0.032***\\
            &     (0.002)   &     (0.002)   &     (0.003)   &     (0.003)   &     (0.003)   &     (0.002)   \\

\multicolumn{7}{l}{\emph{Socio-demographic variables}}\\
Age         &         0.008***&       0.004** &       0.008***&       0.011***&       0.005*  &      -0.003*  \\
            &     (0.001)   &     (0.001)   &     (0.002)   &     (0.003)   &     (0.002)   &     (0.001)   \\
Age$^2$ &     0.000***&       0.000***&       0.000*  &       0.000   &       0.000***&       0.000***\\
            &     (0.000)   &     (0.000)   &     (0.000)   &     (0.000)   &     (0.000)   &     (0.000)   \\
EEA&         -0.055***&      -0.043***&      -0.057***&      -0.062***&      -0.031***&      -0.020***\\
            &     (0.004)   &     (0.004)   &     (0.005)   &     (0.006)   &     (0.007)   &     (0.004)   \\
Non-EEA&     -0.031***&      -0.017***&      -0.008   &      -0.005   &      -0.058***&      -0.031***\\
            &     (0.004)   &     (0.004)   &     (0.006)   &     (0.007)   &     (0.007)   &     (0.005)   \\
In couple  &        0.062***&       0.015***&       0.058***&       0.015***&       0.058***&       0.011***\\
            &     (0.003)   &     (0.002)   &     (0.003)   &     (0.004)   &     (0.004)   &     (0.003)   \\
With dependent children   (yrs<18)   &     0.026***&      -0.032***&       0.023***&      -0.035***&       0.025***&      -0.032***\\
            &     (0.003)   &     (0.003)   &     (0.004)   &     (0.005)   &     (0.005)   &     (0.003)   \\
In couple with dep. children&    0.011** &       0.021***&       0.014** &       0.041***&       0.008   &       0.012** \\
            &     (0.004)   &     (0.003)   &     (0.005)   &     (0.006)   &     (0.007)   &     (0.004)   \\
                        
\multicolumn{7}{l}{\emph{Workplace characteristics}}\\
Part-time  &       -0.097***&      -0.037***&      -0.083***&      -0.034***&      -0.095***&      -0.036***\\
            &     (0.003)   &     (0.002)   &     (0.005)   &     (0.003)   &     (0.004)   &     (0.002)   \\
Public  sector    &      0.031***&       0.059***&     0.003   &       0.005   &       0.059***&       0.088***\\
            &     (0.003)   &     (0.002)   &     (0.005)   &     (0.006)   &     (0.005)   &     (0.003)   \\
 Low gender segregation  &      0.027***&      -0.005   &      -0.141***&      -0.102***&      -0.006   &      -0.012***\\
            &     (0.003)   &     (0.003)   &    (0.009)   &     (0.013)   &     (0.006)   &     (0.003)   \\
Female dominance&  -0.163***&      -0.158***&               &               &               &               \\
            &     (0.003)   &     (0.002)    &               &               &               &               \\
Female Dominance $\times$ &  0.006   &       0.016***&               &               &               &               \\
     \hspace{2mm} Low gender segregation&       (0.005)   &     (0.004)    &               &               &               &               \\
     
\vspace{-3mm}\\

Working region controls&Yes&Yes&Yes&Yes&Yes&Yes\\
Other socio-demographic controls&Yes&Yes&Yes&Yes&Yes&Yes\\
SOC dummies&Yes&Yes&Yes&Yes&Yes&Yes\\
SIC dummies&No&No&Yes&Yes&Yes&Yes\\
Time FE&Yes&Yes&Yes&Yes&Yes&Yes\\
\vspace{-3mm}\\
Observations&        219,027   &      219,297   &      147,953   &       76,699   &       71,074   &      142,598   \\

  \hline\hline
	\end{tabular}	
\begin{tablenotes}[para,flushleft]	
\emph{Notes: Data from UK Labour Force Survey (LFS). Models (1)-(4) are estimated using OLS. Robust errors are in parenthesis. Significance levels: p<0.01 ***, p<0.05 **, p<0.1 *.}
  \end{tablenotes}
  \end{threeparttable}
  }
\end{table}

\clearpage
\section{Figures in Text}

\newpage
\clearpage
\appendix
\clearpage
\newpage

\appendix
\section{Tables in Appendix}
\setcounter{table}{0}
\counterwithin{table}{section}

\begin{table}[th]
\centering
\caption{Share of workers by sector and occupation, male sample}\label{tab:sic_soc_ml}
\scalebox{.75}{
\begin{threeparttable} 
   \begin{tabular}{lcccccccccc}
\vspace{-3mm}\\   
\hline\hline
\vspace{-3mm}\\         
Occupations &1&2&3&4&5&6&7&8&9&   Total\\
\hline
\vspace{-3mm}\\         
\multicolumn{11}{l}{\emph{Sectors}} \\
A - Agriculture, forestry \& fishing&        0.14&        0.03&        0.03&        0.02&        0.28&        0.01&        0.01&        0.10&        0.37&        0.99\\
B - Mining \& quarrying &   0.13&        0.17&        0.11&        0.02&        0.13&        0.00&        0.01&        0.21&        0.02&        0.81\\
C - Manufacturing &       3.03&        2.42&        2.13&        0.53&        4.74&        0.02&        0.28&        4.25&        1.61&       19.01\\
D - Electricity, gas \& air con supply  &0.14&        0.24&        0.15&        0.04&        0.27&        0.00&        0.09&        0.08&        0.03&        1.04\\
E - Water supply, sewerage \& waste&     0.17&        0.14&        0.14&        0.04&        0.11&        0.01&        0.03&        0.44&        0.37&        1.45\\
F - Construction&   1.59&        1.25&        0.64&        0.18&        3.96&        0.02&        0.07&        1.15&        0.95&        9.82\\
G - Distribution&   2.87&        0.54&        1.15&        0.65&        1.94&        0.02&        3.90&        1.69&        2.40&       15.17\\
H - Transport &       0.78&        0.29&        0.63&        0.43&        0.42&        0.23&        0.16&        3.20&        1.71&        7.85\\
I - Accommodation \& food services &        0.70&        0.04&        0.09&        0.13&        1.24&        0.07&        0.20&        0.18&        2.05&        4.70\\
J - Information, communication&         0.83&        2.08&        0.83&        0.14&        0.36&        0.00&        0.21&        0.13&        0.45&        5.04\\
K - Financial \& insurance services&    1.26&        0.92&        1.47&        0.68&        0.04&        0.00&        0.28&        0.02&        0.06&        4.74\\
L - Real estate services&       0.32&        0.13&        0.25&        0.08&        0.11&        0.05&        0.05&        0.02&        0.05&        1.06\\
M - Professional, scientific \& technical activities&    1.33&        2.63&        1.46&        0.49&        0.33&        0.04&        0.14&        0.19&        0.50&        7.11\\
N - Admin \& support services &0.75&        0.60&        0.61&        0.21&        0.48&        0.16&        0.28&        0.25&        1.10&        4.42\\
P - Education &    0.29&        3.96&        0.75&        0.20&        0.17&        0.59&        0.02&        0.07&        0.21&        6.26\\
Q - Health \& social work &     0.66&        1.85&        1.08&        0.35&        0.24&        1.49&        0.06&        0.15&        0.44&        6.32\\
R - Arts, entertainment \& recreation &      0.45&        0.19&        0.59&        0.18&        0.35&        0.28&        0.08&        0.05&        0.34&        2.51\\
S - Other service activities &        0.22&        0.43&        0.21&        0.11&        0.19&        0.21&        0.03&        0.08&        0.14&        1.62\\
T - Households as employers &       0.01&        0.00&        0.00&        0.00&        0.04&        0.02&        0.00&        0.00&        0.01&        0.08\\
\vspace{-3mm}\\         
Total       &          15.68&       17.90&       12.33&        4.48&       15.42&        3.22&        5.89&       12.25&       12.83&      100.00\\

\hline
	\end{tabular}	
\begin{tablenotes}[para,flushleft]	
\emph{Notes: The percentages are calculated on total male employed workers (self-employed are excluded). Labels for occupations: 1 - Managers, Directors And Senior Official ; 2 - Professional Occ.;  3 - Associate Professional And Technical  Occ.; 4 - Administrative And Secretarial Occ.; 5 - Skilled Trades Occ.; 6 - Caring, Leisure And Other Service Occ.;  7 - Sales And Customer Service Occ.; 8 - Process, Plant And Machine Operatives; 9 - Elementary Occ.}
  \end{tablenotes}
  \end{threeparttable}
  }
\end{table}

\begin{table}[th]
\centering
\caption{Share of workers by sector and occupation, female sample}\label{tab:sic_soc_fml}
\scalebox{.75}{
\begin{threeparttable} 
   \begin{tabular}{lcccccccccc}
\vspace{-3mm}\\   
\hline\hline
\vspace{-3mm}\\         
Occupations &1&2&3&4&5&6&7&8&9&   Total\\
\hline
\vspace{-3mm}\\         
\multicolumn{11}{l}{\emph{Sectors}} \\
A - Agriculture, forestry \& fishing &        0.04&        0.02&        0.02&        0.09&        0.04&        0.05&        0.02&        0.02&        0.11&        0.41\\
B - Mining \& quarrying  &        0.03&        0.03&        0.04&        0.06&        0.00&        0.00&        0.00&        0.00&        0.00&        0.17\\
C - Manufacturing  &        0.79&        0.52&        0.98&        1.61&        0.27&        0.04&        0.34&        1.17&        0.66&        6.38\\
D - Electricity, gas \& air con supply  &        0.04&        0.05&        0.07&        0.10&        0.00&        0.00&        0.11&        0.00&        0.01&        0.38\\
E - Water supply, sewerage \& waste  &        0.04&        0.04&        0.06&        0.12&        0.00&        0.00&        0.04&        0.01&        0.03&        0.35\\
F - Construction &        0.27&        0.19&        0.26&        0.98&        0.04&        0.01&        0.09&        0.01&        0.06&        1.91\\
G - Distribution&        1.51&        0.34&        1.05&        2.00&        0.25&        0.07&        8.71&        0.25&        1.41&       15.59\\
H - Transport  &        0.25&        0.09&        0.21&        0.67&        0.01&        0.41&        0.20&        0.17&        0.43&        2.44\\
I - Accommodation \& food services  &        0.58&        0.03&        0.16&        0.41&        0.49&        0.27&        0.40&        0.03&        3.70&        6.09\\
J - Information, communication&        0.31&        0.50&        0.47&        0.40&        0.02&        0.01&        0.27&        0.01&        0.12&        2.11\\
K - Financial \& insurance services&        0.70&        0.44&        1.05&        1.95&        0.00&        0.01&        0.51&        0.01&        0.05&        4.72\\
L - Real estate services &        0.26&        0.07&        0.40&        0.45&        0.01&        0.05&        0.11&        0.00&        0.04&        1.39\\
M - Professional, scientific \& technical activities &        0.78&        1.38&        1.29&        2.02&        0.03&        0.18&        0.22&        0.04&        0.35&        6.28\\
N - Admin \& support services&        0.36&        0.21&        0.47&        0.67&        0.03&        0.35&        0.35&        0.03&        0.95&        3.41\\
P - Education &        0.39&        8.30&        1.14&        1.71&        0.20&        4.69&        0.06&        0.01&        1.76&       18.27\\
Q - Health \& social work&        1.60&        5.86&        4.17&        3.16&        0.20&        8.65&        0.20&        0.04&        1.00&       24.87\\
R - Arts, entertainment \& recreation&        0.31&        0.21&        0.44&        0.61&        0.04&        0.28&        0.15&        0.01&        0.39&        2.43\\
S - Other service activities &        0.20&        0.25&        0.25&        0.44&        0.03&        1.07&        0.09&        0.02&        0.22&        2.58\\
T - Households as employers  &        0.00&        0.00&        0.00&        0.01&        0.00&        0.17&        0.00&        0.00&        0.04&        0.22\\
\vspace{-3mm}\\         
Total       &        8.46&       18.51&       12.54&       17.46&        1.68&       16.31&       11.86&        1.85&       11.34&      100.00\\
\hline
	\end{tabular}	
\begin{tablenotes}[para,flushleft]	
\emph{Notes: The percentages are calculated on total female employed workers (self-employed are excluded). Labels for occupations: 1 -- Managers, Directors And Senior Official ; 2 -- Professional Occ.;  3 -- Associate Professional And Technical  Occ.; 4 -- Administrative And Secretarial Occ.; 5 -- Skilled Trades Occ.; 6 -- Caring, Leisure And Other Service Occ.;  7 -- Sales And Customer Service Occ.; 8 -- Process, Plant And Machine Operatives; 9 -- Elementary Occ.}
  \end{tablenotes}
  \end{threeparttable}
  }
\end{table}

\begin{table}[!ht]
    \centering
       \caption{Post-estimation OLS coefficients of  variables selected by  \textsc{Lasso}}\label{tab:lasso} 
    \scalebox{.65}{
    \begin{threeparttable} 	
\begin{tabular}{lcccccc}
    \hline\hline
            \vspace{-3mm}\\
	&	Log(wage)	&	Permanent job	&	Part-time job	&	Log(hours)	&	Remote work 	&	Female-dom. sector	\\
    &	(1)	&	(2)	&	(3)&	(4)	&	(5) 	& (6)	\\
        \hline
        \vspace{-3mm}\\
Female	&	-0.006	&&	0.048	&-0.081&&	0.095	\\
In couple	&&	0.008&&&	0.0112	&		\\
With dep. Children	(yrs<18)	&	0.004	&&&&	0.001	&		\\
Female x In couple	&		&		&		&		&		&		\\
\hline
    \end{tabular}
    \begin{tablenotes}[para,flushleft]	
\emph{Note: The table reports the post-estimation OLS coefficients of the variables selected by Lasso. The estimated models correspond to those with minimum BIC. The \textsc{Stata} command used to obtain the estimates is \texttt{lasso2}.}
  \end{tablenotes}
  \end{threeparttable}
    }
\end{table}
 \begin{table}[th!]
\centering
\caption{Probit for female dominated sectors}\label{tab:probit_fmldom}
\scalebox{.8}{
\begin{threeparttable} 	
\begin{tabular}{llll}
\hline\hline
\vspace{-3mm}\\
&    Full sample      &    Male sample & Female sample       \\
\hline
\vspace{-3mm}\\
\multicolumn{4}{l}{\emph{Dep. var: Female dominated sectors}}\\
\vspace{-3mm}\\
Woman      &     1.063***&               &               \\
            &  (0.007)    &               &               \\
Woman not in couple&     -0.218***&               &               \\
\hline
\end{tabular}
\begin{tablenotes}[para,flushleft]	
\emph{Notes: The estimates are used to calculate propensity scores for the PSM. Robust standard errors. Significance levels: pvalue<0.01 ***, pvalue<0.05 **, pvalue<0.1 *.}
  \end{tablenotes}
  \end{threeparttable}
 }
\end{table}

\newgeometry{left=10mm,right=10mm,top=2.5cm,bottom=2cm}

\begin{table}[ht]
\centering
\caption{Contribution of individual components of KBO decomposition, female dominated sectors}\label{tab:oaxaca_factors_fmldom}
\scalebox{.41}{
\begin{threeparttable} 	
   \begin{tabular}{lccccccccccccccc}    
\vspace{-3mm}\\   
\hline\hline
\vspace{-3mm}\\
 
 Year           &  2006   &  2007   &  2008   &  2009   &  2010   &  2011   &  2012   &  2013   &  2014   &  2015   &  2016   &  2017   &  2018   &  2019   &  2020   \\
 \vspace{-3mm}\\
 \hline
  \vspace{-2mm}\\
\emph{Endowment effect}          &               &               &               &               &               &               &               &               &               &               &               &               &               &               &               \\
Low gender sect. segregation&        0.01***&        0.01** &        0.01***&        0.02***&        0.01** &        0.01***&        0.02***&        0.02***&        0.01***&        0.01***&        0.00   &        0.01***&        0.01***&        0.01***&        0.03***\\
                    &     (0.003)   &     (0.003)   &     (0.003)   &     (0.003)   &     (0.003)   &     (0.003)   &     (0.004)   &     (0.004)   &     (0.003)   &     (0.003)   &     (0.003)   &     (0.004)   &     (0.004)   &     (0.004)   &     (0.005)   \\
\hline
	\end{tabular} 
\begin{tablenotes}[para,flushleft]	
\emph{Notes: Contribution of main socio-demographic characteristics, human capital attributes and sectoral indicators. Occupations (SOC) are omitted for layout reasons. Significance levels: pvalue<0.01 ***, pvalue<0.05 **, pvalue<0.1 *.}
  \end{tablenotes}
  \end{threeparttable}

  }
\end{table}

\begin{table}[ht]
\centering
\caption{Contribution of individual components of KBO decomposition, male dominated sectors}\label{tab:oaxaca_factors_mldom}
\scalebox{.41}{
\begin{threeparttable} 	
   \begin{tabular}{lccccccccccccccc}    
\vspace{-3mm}\\   
\hline\hline
\vspace{-3mm}\\
 
 Year           &  2006   &  2007   &  2008   &  2009   &  2010   &  2011   &  2012   &  2013   &  2014   &  2015   &  2016   &  2017   &  2018   &  2019   &  2020   \\
 \vspace{-3mm}\\
 \hline
  \vspace{-2mm}\\
\emph{Endowment effect}    &               &               &               &               &               &               &               &               &               &               &               &               &               &               &               \\
Low gender sect. segregation&        0.05***&        0.04***&        0.05***&        0.06***&        0.05***&        0.04***&        0.05***&        0.04***&        0.03***&        0.03***&        0.02** &        0.02***&        0.03***&        0.04***&        0.05***\\
                    &     (0.011)   &     (0.009)   &     (0.010)   &     (0.013)   &     (0.012)   &     (0.014)   &     (0.011)   &     (0.011)   &     (0.009)   &     (0.009)   &     (0.011)   &     (0.009)   &     (0.010)   &     (0.009)   &     (0.010)   \\
\hline
	\end{tabular} 
\begin{tablenotes}[para,flushleft]	
\emph{Notes: Contribution of main socio-demographic characteristics, human capital attributes and sectoral indicators. Occupations (SOC) are omitted for layout reasons. Significance levels: pvalue<0.01 ***, pvalue<0.05 **, pvalue<0.1 *.}
  \end{tablenotes}
  \end{threeparttable}

  }
\end{table}

\restoregeometry

\begin{table}[ht]
\caption{Kolmogorov--Smirnov test for equality of distributions}\label{tab:ksmirnov}
\scalebox{.55}{
\begin{threeparttable} 	
\centering
   \begin{tabular}{lcccc}
\hline\hline
\vspace{-1mm}\\
 &	\shortstack{Men vs Women \\in female-dominated sectors}	&	\shortstack{Men vs Women	\\in male-dominated sectors}	&	\shortstack{Workers in female-dominated sectors \\  vs  male-dominated sectors if women	}&	\shortstack{Workers in female-dominated sectors  \\  vs male-dominated sectors if men	} \\ 
& (1)&(2)&(3)&(4)\\
\hline
\vspace{-1mm}\\

Predicted wage&	0.1700  	&	0.1541 	&	0.2438 	&	0.1890  	\\
Residual wage	&	0.0366 	&	0.0243  	&	0.0442 	&	0.0231    	\\
Counterfactual predicted wage	&	0.1055	&	0.0871&	0.0867	&	0.0471	\\
Counterfactual residual wage	&	0.0821  	&	0.1136 	&	0.1518  	&	0.1791  	\\
  \hline\hline
	\end{tabular}	
\begin{tablenotes}[para,flushleft]	
\emph{Notes: The Kolmogorov-Smirnov test is a non-parametric test for comparing the equality of two distributions without imposing the normality assumption. The first two columns compare the distribution of men against the distribution of women working in female-dominated sectors (Column 1) or male-dominated sectors (Column 2).  The last two columns compare the distribution of workers in female-dominated sectors against those in male-dominated sectors if women (Column 3) or if men (Column 4). All test statistics are significant at $1\%$ level.}
  \end{tablenotes}
  \end{threeparttable}
  }
\end{table}

\clearpage
\newpage
\section{Figures in Appendix}

\setcounter{figure}{0}
\counterwithin{figure}{section}

\begin{figure}[th!] 
	 \caption{Covariate imbalance test, single components}\label{fig:balance}
  \centering
	\subfloat[Full sample]{\includegraphics[scale=.55]{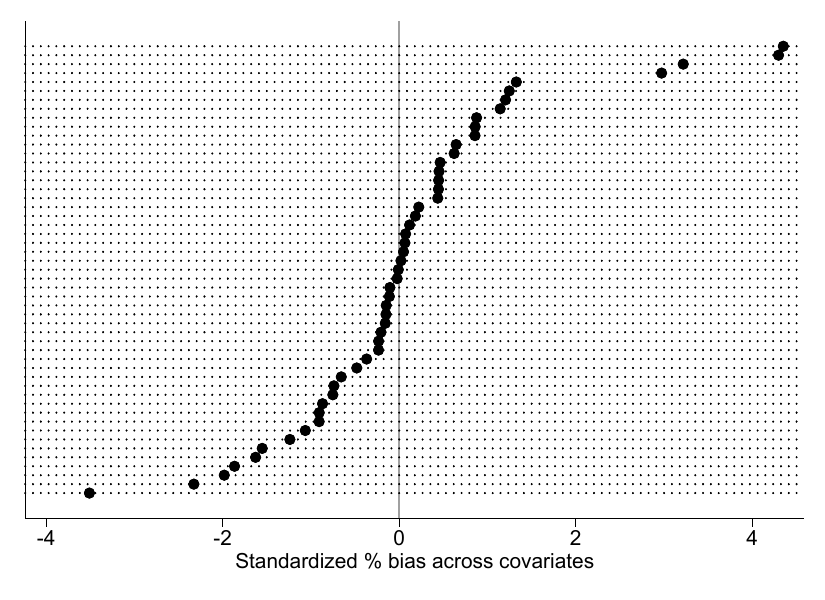}}\\
	\subfloat[Male sample]{\includegraphics[scale=.55]{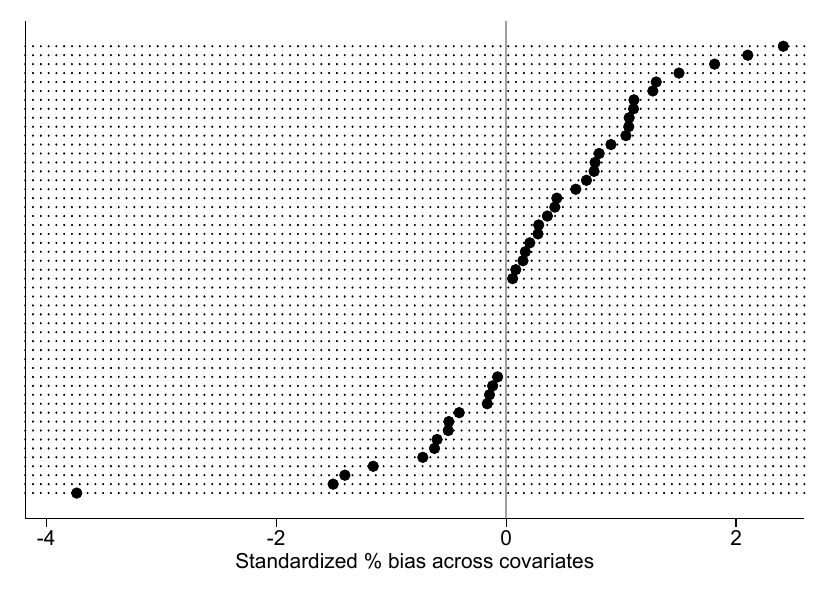}}\\
	\subfloat[Female sample]{\includegraphics[scale=.55]{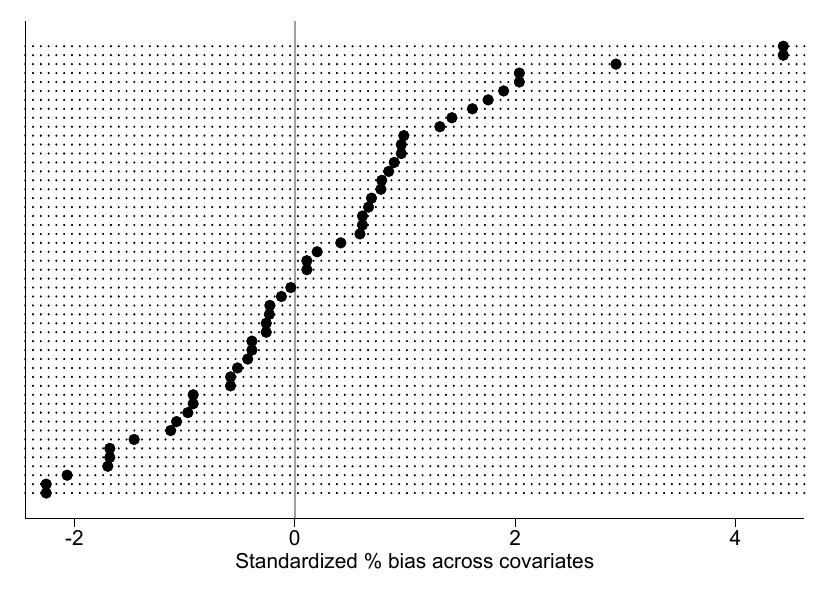}}\\
	\begin{minipage}{.95\linewidth}
	\vspace{3mm}
	\footnotesize
	\emph{Note: The included covariates are balanced if the standardised bias after matching is within $\pm 5\%$ \citep{rosenbaum1985}. If the condition is satisfied, the matching method successfully builds a valid control group.}
	\end{minipage}
\end{figure}
\clearpage
\section{Additional material: Tables}\label{sec:additional_tab}

\setcounter{table}{0}
\counterwithin{table}{section}

\begin{table}[th]
\centering
\caption{Share of workers by sector and occupation}\label{tab:sic_soc_tot}
\scalebox{.75}{
\begin{threeparttable} 
   \begin{tabular}{lcccccccccc}
\vspace{-3mm}\\   
\hline\hline
\vspace{-3mm}\\         
Occupations &1&2&3&4&5&6&7&8&9&   Total\\
\hline
\vspace{-3mm}\\         
\multicolumn{11}{l}{\emph{Sectors}} \\
A - Agriculture, forestry \& fishing&        0.09&        0.02&        0.03&        0.05&        0.16&        0.03&        0.01&        0.06&        0.24&        0.69\\
B - Mining \& quarrying   &        0.08&        0.10&        0.08&        0.04&        0.07&        0.00&        0.01&        0.10&        0.01&        0.48\\
C - Manufacturing   &        1.89&        1.45&        1.55&        1.08&        2.46&        0.03&        0.31&        2.68&        1.13&       12.57\\
D - Electricity, gas \& air con supply&        0.09&        0.14&        0.11&        0.07&        0.13&        0.00&        0.10&        0.04&        0.02&        0.70\\
E - Water supply, sewerage \& waste &        0.11&        0.09&        0.10&        0.08&        0.05&        0.00&        0.03&        0.22&        0.20&        0.89\\
F - Construction&        0.92&        0.71&        0.45&        0.59&        1.96&        0.02&        0.08&        0.57&        0.49&        5.78\\
G - Distribution &        2.17&        0.44&        1.10&        1.34&        1.08&        0.05&        6.36&        0.95&        1.90&       15.38\\
H - Transport   &        0.51&        0.19&        0.42&        0.55&        0.21&        0.32&        0.18&        1.65&        1.06&        5.08\\
I - Accommodation \& food services  &        0.64&        0.03&        0.13&        0.27&        0.86&        0.17&        0.30&        0.11&        2.90&        5.41\\
J - Information, communication &        0.56&        1.27&        0.65&        0.27&        0.19&        0.00&        0.24&        0.07&        0.28&        3.54\\
K - Financial \& insurance services &        0.98&        0.67&        1.25&        1.33&        0.02&        0.01&        0.40&        0.01&        0.06&        4.73\\
L - Real estate services &        0.29&        0.10&        0.33&        0.27&        0.06&        0.05&        0.08&        0.01&        0.05&        1.23\\
M - Professional, scientific \& technical activities &        1.05&        1.99&        1.37&        1.27&        0.18&        0.11&        0.18&        0.11&        0.42&        6.69\\
N - Admin \& support services&        0.55&        0.40&        0.54&        0.44&        0.25&        0.25&        0.31&        0.14&        1.02&        3.90\\
P - Education&        0.34&        6.18&        0.95&        0.97&        0.19&        2.68&        0.04&        0.04&        1.00&       12.39\\
Q - Health \& social work &        1.14&        3.90&        2.66&        1.78&        0.22&        5.15&        0.13&        0.09&        0.72&       15.79\\
R - Arts, entertainment \& recreation&        0.38&        0.20&        0.51&        0.40&        0.19&        0.28&        0.12&        0.03&        0.36&        2.47\\
S - Other service activities&        0.21&        0.34&        0.23&        0.28&        0.11&        0.65&        0.06&        0.05&        0.18&        2.11\\
T - Households as employers &        0.00&        0.00&        0.00&        0.00&        0.02&        0.09&        0.00&        0.00&        0.03&        0.15\\
\vspace{-3mm}\\         
Total       &       11.99&       18.21&       12.44&       11.11&        8.41&        9.90&        8.94&        6.94&       12.07&      100.00\\
\hline
	\end{tabular}	
\begin{tablenotes}[para,flushleft]	
\emph{Notes: The percentages are calculated on total employed workers (self-employed are excluded). Labels for occupations:  1 - Managers, Directors And Senior Official ; 2 - Professional Occ.;  3 - Associate Professional And Technical  Occ.; 4 - Administrative And Secretarial Occ.; 5 - Skilled Trades Occ.; 6 - Caring, Leisure And Other Service Occ.;  7 - Sales And Customer Service Occ.; 8 - Process, Plant And Machine Operatives; 9 - Elementary Occ.}
  \end{tablenotes}
  \end{threeparttable}
  }
\end{table}

\newgeometry{left=10mm,right=10mm,top=2.5cm,bottom=2cm}

\begin{table}[ht]
\centering
\caption{Contribution of individual components of KBO decomposition, full sample}\label{tab:oaxaca_factors}
\scalebox{.5}{
\begin{threeparttable} 	
\begin{tabular}{lccccccccccccccc}    
\vspace{-3mm}\\   
\hline\hline
\vspace{-3mm}\\
 Year           &  2006   &  2007   &  2008   &  2009   &  2010   &  2011   &  2012   &  2013   &  2014   &  2015   &  2016   &  2017   &  2018   &  2019   &  2020   \\
 \hline
 
 \vspace{-1mm}
\emph{Endowment effect} &               &               &               &               &               &               &               &               &               &               &               &               &               &               &               \\
Log(hours)          &       -0.11***&       -0.08***&       -0.15***&       -0.11***&       -0.12***&       -0.12***&       -0.11***&       -0.06***&       -0.10***&       -0.16***&       -0.12***&       -0.11***&       -0.12***&       -0.08***&       -0.07***\\
                    &     (0.013)   &     (0.013)   &     (0.013)   &     (0.013)   &     (0.013)   &     (0.013)   &     (0.013)   &     (0.013)   &     (0.014)   &     (0.014)   &     (0.014)   &     (0.014)   &     (0.015)   &     (0.015)   &     (0.015)   \\
Log(hours)$^2$      &        0.11***&        0.07***&        0.15***&        0.11***&        0.12***&        0.12***&        0.10***&        0.06***&        0.10***&        0.16***&        0.12***&        0.10***&        0.12***&        0.07***&        0.06***\\
                    &     (0.014)   &     (0.014)   &     (0.015)   &     (0.014)   &     (0.015)   &     (0.015)   &     (0.015)   &     (0.015)   &     (0.015)   &     (0.015)   &     (0.015)   &     (0.015)   &     (0.016)   &     (0.016)   &     (0.016)   \\
\hline
	\end{tabular} 
\begin{tablenotes}[para,flushleft]	
\emph{Notes: Contribution of main socio-demographic characteristics, human capital attributes and sectoral indicators. Significance levels: pvalue<0.01 ***, pvalue<0.05 **, pvalue<0.1 *.}
  \end{tablenotes}
  \end{threeparttable}

  }
\end{table}	
\restoregeometry

\clearpage
\newpage
\section{Additional material: Figures}
\setcounter{figure}{0}
\counterwithin{figure}{section}

\begin{figure}[h!] 
	 \caption{KBO decomposition, full sample}\label{fig:oaxaca}
  \centering
	\includegraphics[scale=.8]{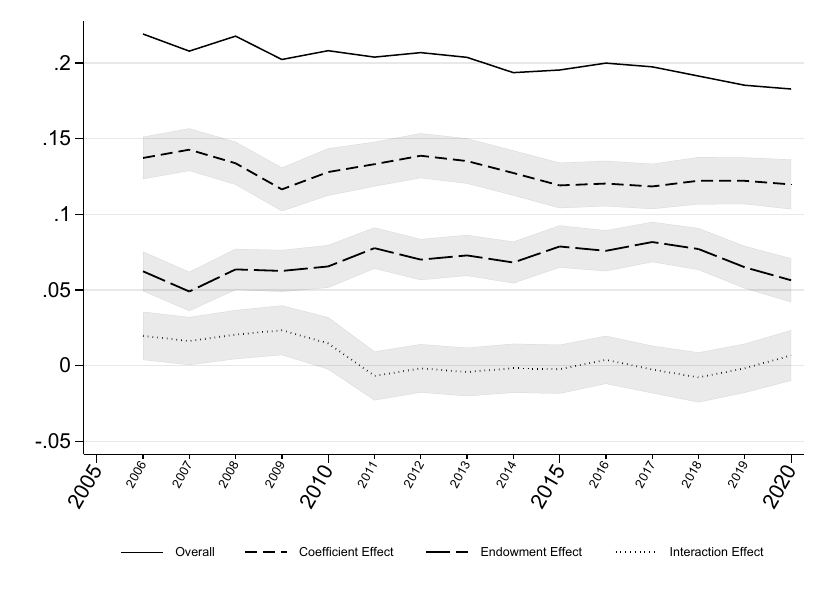}\\
	 \begin{minipage}{.95\linewidth}
	\vspace{3mm}
	\footnotesize
	\emph{\textsc{Note}: The figure shows the evolution of the components of the KBO decomposition and their sum over time for full sample. Women are contrasted to men. The dashed line represents the \emph{coefficient effect}, the long-dashed line the \emph{endowment effect} and the dotted line the part of the ``unexplained'' component of the three-fold decomposition (or \emph{interaction effect}). The corresponding shadowed areas display the 95\% confidence intervals. The solid line is the sum of the three effects and reveals their overall contribution. The wage difference between men and women is, on average, 0.2 logarithmic points over time. Most of the gender pay gap (around three-fourths) can be explained by differences in the estimated coefficients between genders. The \emph{coefficient effect} on average quantifies indeed an increase of 0.16 logarithmic points in women's wages when the male coefficients are applied to female characteristics. In addition, this component displays a downward trend after 2008.  The \emph{endowment effect} would quantify an expected average increase in women's wage by around 0.05 points if they had male predictors levels. Therefore, differences in observed characteristics account for one-fourth of the gap.\\
\textsc{Estimation note}: Both models for women and men are estimated using the Mincerian regression equation (with OLS). The shaded areas is the 95\% confidence intervals.}
	\end{minipage}
\end{figure}

\end{document}